\documentclass{aa}  
\usepackage{graphicx}
\usepackage{txfonts}
\usepackage{multirow}
\usepackage{hyperref}
\usepackage{placeins}
\usepackage[usenames,dvipsnames]{color}
\usepackage{natbib}
\bibpunct{(}{)}{;}{a}{}{,} 

\begin{document} 

   \title{MaNGA AGN dwarf galaxies (MAD)}
   \subtitle{III. The role of mergers and environment in active galactic nucleus activity in dwarf galaxies}

   \author{
    A. Er\'{o}stegui\inst{1,}\thanks{\email{erostegui@ice.csic.es}},
    M. Mezcua\inst{1,2},
    M. Siudek\inst{1,3},
    H. Dom\'{i}nguez S\'{a}nchez\inst{4},
    V. Rodr\'{i}guez Morales\inst{1}
    }
      
   \institute{Institute of Space Sciences (ICE, CSIC), Campus UAB, Carrer de Magrans, s/n, 08193 Barcelona, Spain
         \and
         Institut d’Estudis Espacials de Catalunya (IEEC), Edifici RDIT, Campus UPC, 08860 Castelldefels (Barcelona), Spain
         \and
             Instituto de Astrof\'isica de Canarias, Av. V\'ia L\'actea, s/n, E38205 La Laguna, Spain
        \and
            Instituto de F\'isica de Cantabria (IFCA, CSIC), Avda. de los Castros, s/n, 39005 Santander, Spain
            }

   \date{Received MM DD, YY; accepted MM DD, YY}
 
  \abstract
   {Investigating whether and how galaxy mergers affect black hole growth can be a determinant for black hole-galaxy coevolution models and, in particular, for understanding how early Universe seed black holes grew to become supermassive. However, while mergers have been observed to enhance the active galactic nucleus (AGN) activity and thus black hole growth in massive galaxies, it is not yet known how this relation and the role of the environment translates to dwarf galaxies (the most likely hosts of the early seed black holes) since there are scarce and mixed results in the literature.}
   {We seek to assess the impact of galaxy mergers and the environment on AGN triggering in dwarf galaxies.}
   {We used a sample of 3280 dwarf galaxies with integral-field spectroscopic data from the MaNGA survey to study the AGN fraction throughout the merger process and how it is affected by the environment (characterized by galaxy isolation, being in a void, and group richness). We also compare the fraction of interacting galaxies in AGN and non-AGN dwarf galaxies.}
   {We find that dwarf galaxy mergers can ignite AGNs at separations below 20 kpc. The AGN fraction increases notoriously after the first pass and remains enhanced until the final stage. However, mergers are not the dominant AGN triggering mechanism. We also find that the environment has a non-negligible impact on AGN activity in dwarf galaxies, as the AGN fraction increases when moving to lower density environments. These findings provide the most statistically robust constraints to date on the effects of dwarf galaxy mergers and environment on AGN activity and black hole growth.}
   {}

    \titlerunning{MAD - III. The role of mergers and environment in AGN in dwarf galaxies}
    \authorrunning{A. Er\'{o}stegui et al.}
   \maketitle
   
\section{Introduction} \label{sec:intro} 
A detailed model of the growth history of supermassive black holes (SMBHs; $\geq 10^{6} \text{ M}_{\odot}$) can be a determinant to unveiling their origin. The detection of SMBHs at redshifts beyond $z = 10$ \citep{Maiolino2024} suggests they had to start out as seed intermediate-mass black holes (IMBHs; $10^{2} < M_{\text{BH}} \lesssim 10^{6} \text{ M}_{\odot}$) at even higher redshifts. Massive black holes (BHs; i.e., SMBHs and IMBHs) grow mainly via accretion \citep{Volonteri2010} observed as active galactic nuclei (AGNs), and coalescence with other massive BHs \citep[e.g.,][]{Goulding2019}, which can be detected via gravitational waves or, at times earlier than the coalescence, as a binary or dual AGN. Although BH coalescence is evidently consequential to a galaxy merger, it is also believed that AGN activity is ignited or enhanced during a galactic merger, as this process provides the gas necessary to build the central stellar mass and feed the SMBH \citep{Hopkins2006, Wild2007}. In fact, in the massive galaxy regime, the presence of AGNs has been observed to be related to interacting galaxies (e.g., \citealp{Ellison2011, Ellison2013, Satyapal2014, Steffen2023, Barrows2023, Li2023}), although there are seemingly contrasting results on this \citep{Villforth2023}. The role of interaction in triggering AGNs is even more unclear in the dwarf galaxy regime (stellar mass $M_{\star} \leq 10^{10} \text{ M}_{\odot}$)\footnote{While there is not a consensus on the upper mass limit of dwarf galaxies, this value has been adopted by several authors \citep[e.g.,][]{Paudel2018, Manzano2020, Bichanga2024}.} since very few studies have investigated this relation while focusing on low stellar masses, and so far the results are mixed \citep{Kaviraj2019, Bichanga2024, Micic2024}. \par
Having knowledge on how galaxy mergers affect AGN activity at different mass scales is of special importance to constructing a precise model of BH growth history and more so for dwarf galaxies since they are the dominant population at all redshifts \citep[e.g.,][]{Martin2019, Driver2022}, and according to the hierarchical model of galaxy formation, dwarf galaxies give origin to massive galaxies and their respective SMBHs. Moreover, local dwarf galaxies are thought to have a quieter merger history than primordial galaxies in the same mass range, and consequently the mass of their IMBHs is expected to not have grown significantly through BH coalescence and accretion, so they could still hold some dependence on the initial seed BH mass \citep{Wassenhove2010}. However, dwarf galaxies undergo on average three major mergers after their formation \citep{Fakhouri2010}, each of which could lead to the coalescence of the central BHs and trigger rapid BH accretion, resulting in significant BH growth \citep{Deason2014}. In this case, the IMBHs in local dwarf galaxies should not be treated as relics of the early Universe seed BHs \citep{Mezcua2019a}. It is not yet clear if BH coalescence and rapid BH accretion occur in dwarf galaxies. In the case of the former, the dynamical friction of dwarf galaxies might not be efficient enough to remove the necessary angular momentum to form a close BH binary, and consequently the BHs might not merge \citep{Tamfal2018}. As for the latter, no binary nor dual AGN system has been confirmed yet in dwarf galaxies, and the effect (or lack thereof) of dwarf galaxy mergers on AGN triggering remains unestablished. This very last point is the main problem we address in this work. \par
Another important aspect to take into account is the impact of the environment on AGN activity. The presence of AGNs in different environments has been studied by several authors, and they have found higher AGN fractions for massive galaxies in the field and lower density regions \citep[e.g.,][]{Kauffmann2004, Silverman2009, Sabater2013, Ehlert2014, Lopes2017}, although others found no to very weak environmental dependence \citep[e.g.,][]{Martini2007, Pimbblet2013, Man2019}, even for the dwarf galaxy regime \citep{Amiri2019, Siudek2022}. Settling what environmental conditions (if any) affect AGN triggering in dwarf galaxies is also necessary to constraining the BH growth history model. In this work, we investigate how the AGN fraction of dwarf galaxies evolves throughout different stages of the galaxy merger process with unprecedented statistical robustness and how it is affected by three environmental factors: isolation, voids, and group richness. We also investigate the extent to which galaxy mergers can account for the AGN population in dwarf galaxies. The sample, its classification, and processing are described in Sect.~\ref{sec:data_class}, the results are reported in Sect.~\ref{sec:Results}, and conclusions are presented in Sect.~\ref{sec:Conclusions}. Throughout the paper, we assume a $\Lambda \text{-CDM}$ cosmology with parameters $H_{0} = 73~\text{km~s}^{-1}~\text{Mpc}^{-1}$, $\Omega_{\Lambda} = 0.73$, and $\Omega_{\text{m}} = 0.27$.

\section{Data and classification} 
\label{sec:data_class}
The Mapping Nearby Galaxies at Apache Point Observatory (MaNGA) survey is an integral-field spectroscopy survey carried out with the Sloan Digital Sky Survey (SDSS) 2.5 m telescope \citep{Gunn2006} as part of SDSS-IV \citep{Blanton2017}. The final data release \citep[DR17;][]{Abdurrouf2022} contains all the observations and science data products for the complete survey, which includes over 10,000 galaxies with redshifts up to $z=0.15$ that have been selected from the NASA-Sloan Atlas \citep[NSA\footnote{\url{http://www.nsatlas.org}};][]{Wake2017}. MaNGA makes use of 17 integral-field units (IFU) that vary in diameter from 12\arcsec to 32\arcsec (19 to 127 fibers with 2\arcsec of diameter) and has a wavelength coverage of 3600-10300 \r{A} with a spectral resolution of $R\sim2000$ and instrumental velocity dispersion of $70~\text{km~s}^{-1}$ \citep[1$\sigma$;][]{Law2021, Yan2016}.

The MaNGA survey comprises three subsamples: the Primary, Secondary, and Color-Enhanced samples. The Primary and Secondary samples were designed to achieve a flat number density with respect to the i-band absolute magnitude $M_{i}$ and a spectroscopic coverage up to 1.5 and 2.5 effective radii ($R_{e}$), respectively. The Color-Enhanced sample was selected to increase the number of galaxies in underpopulated regions of the color $(\text{NUV}-i)$ vs. absolute magnitude ($M_{i}$) plane of the Primary sample. The combination of the Primary and Color-Enhanced samples is referred to as the Primary+ sample \citep{Wake2017}.

\begin{figure}
    \includegraphics[width=\columnwidth]{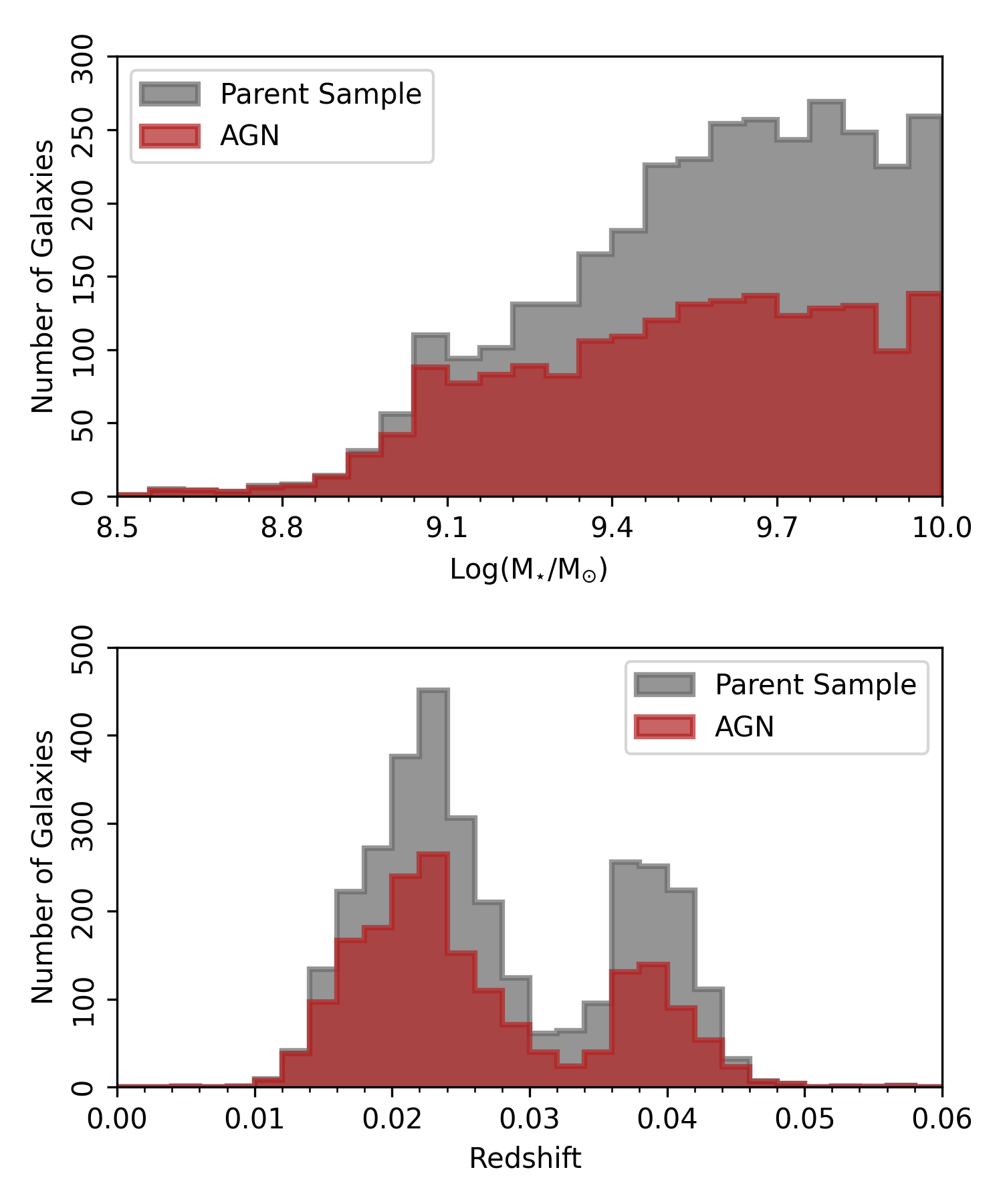}
    \caption{Distribution of the NSA Petrosian stellar masses (top) and redshift (bottom) of the final sample of 3280 unique dwarf galaxies and the subset of galaxies that are classified as AGN (see Sect.~\ref{sec:AGNsel}).}
    \label{fig:mass_z_dist}
\end{figure}

The parent sample used in this paper is detailed in \citet{Mezcua2024}, and it consists of 3306 galaxies drawn from the MaNGA DR17 full sample by considering only those galaxies with Petrosian and Sérsic NSA stellar masses consistent within 0.5 dex \citep{Secrest2020}, then selecting galaxies with stellar mass $M_{\star}\le10^{10}~\text{M}_{\odot}$, and finally removing duplicates according to the \texttt{DUPL\_GR} parameter from the MaNGA Deep Learning Morphology catalog \citep{Dominguez2022}. For this work we removed from the sample those galaxies with $z>0.06$ (4) and $z<0.005$ (11), as they are too few and sparsely sampled, so they cannot be properly matched with the control sample. We also removed another 11 galaxies due to different issues found during the inspection (e.g., wrong NSA redshifts, not being a galaxy, MaNGA field centered on a different object). The final sample is composed of 3280 MaNGA dwarf galaxies. Figure~\ref{fig:mass_z_dist} shows the stellar mass and redshift distributions of the main sample. We note that the bimodality of the redshift distribution is due to the dwarf galaxy stellar mass cut combined with the design of the MaNGA survey. As a result, the first peak (lower redshift) corresponds to galaxies from the Primary+ sample, while the second peak (higher redshift) corresponds mostly to galaxies from the Secondary sample.

\subsection{AGN selection} 
\label{sec:AGNsel}
Here we give a brief summary of the AGN classification of our sample, which is detailed in \citet{Mezcua2024}. The IFU spaxels were classified based on the [NII]-BPT, [SII]-BPT, and [OI]-BPT optical emission line diagnostic diagrams with the \citet{Kewley2001,Kewley2006} and \citet{Kauffmann2003} demarcation lines to differentiate between AGN, star formation (SF), and composite (mixture of AGN and SF) ionization in the [NII]-BPT and between Seyfert, SF, and LINER\footnote{Low ionization nuclear emission line region.} ionization in the [SII]-BPT and [OI]-BPT. Additionally, the WHAN diagram \citep{CidFernandes2010} was used to differentiate between true AGN ($\text{H}\alpha$ equivalent width $\text{(EW)}\ge 3$ \r{A} and the flux ratio log [NII]/$\text{H}\alpha\ge~-0.4$) and SF. Based on the combination of the outcomes of all diagrams, spaxels were then classified as AGN, AGN-WHAN, composite, SF-AGN,\footnote{These are spaxels that were classified as SF in the [NII]-BPT but as Seyfert in the [SII]- or [OI]-BPT, unlike composite spaxels, which are classified as such in the [NII]-BPT, and as SF or LINER in the [SII]- or [OI]-BPT.} LINER, or SF.
Then, spaxels with a signal-to-noise ratio of $S/N\geq3$ and unmasked stellar velocity and stellar velocity dispersion were considered as valid, and if the ratio of the class to valid spaxels was 5\% or higher, the galaxies were classified as AGN, AGN-WHAN, SF-AGN, composite, or LINER.

\begin{figure}
    \includegraphics[width=\columnwidth]{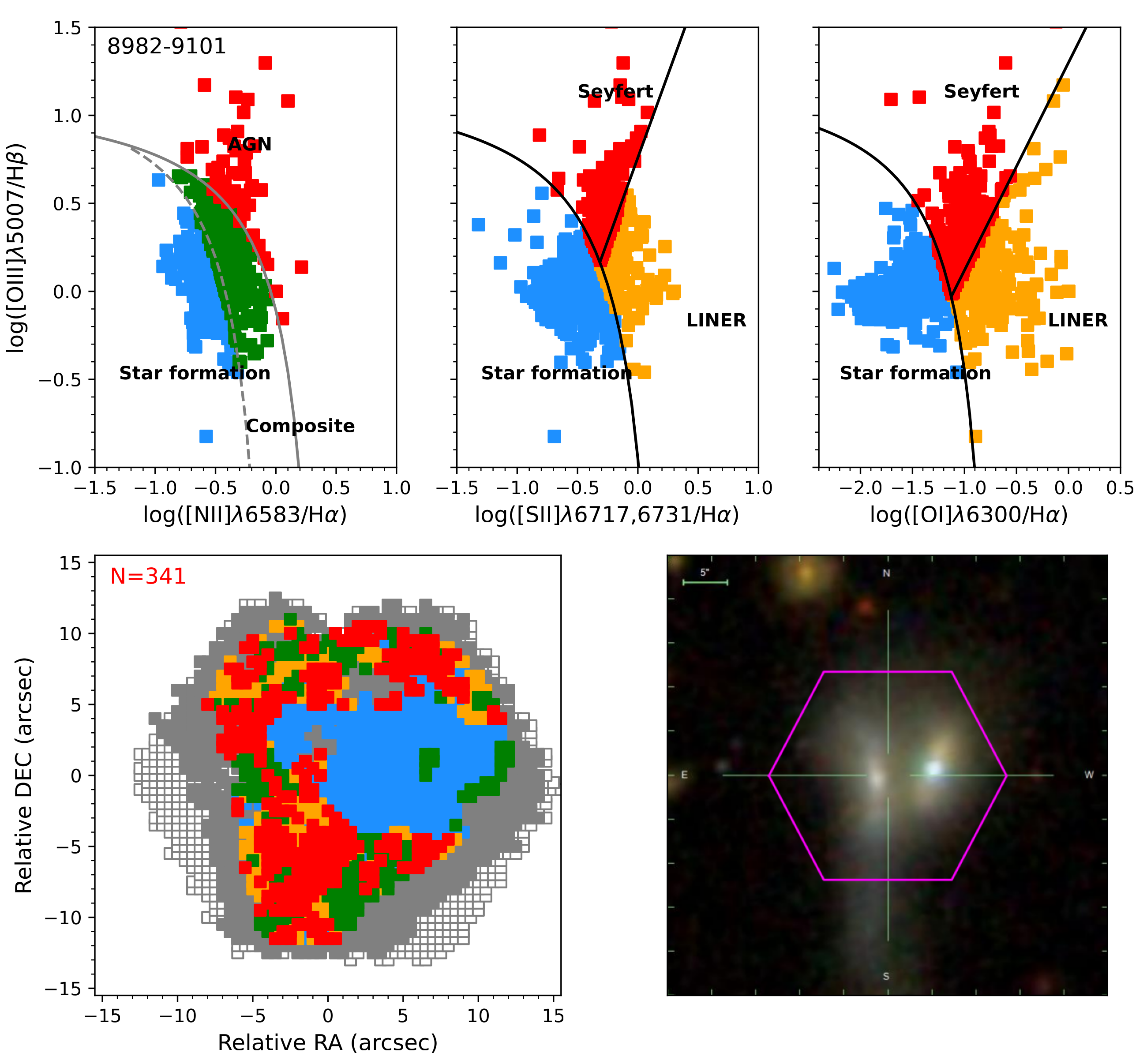}
    \caption{Analysis of MaNGA data of the AGN dwarf galaxy 8982-9101. Top row: Location of each MaNGA spaxel on the [NII]-BPT (left), [SII]-BPT (middle), and [OI]-BPT (right) used to distinguish between ionization by AGN (red spaxels), SF (blue spaxels), composite (green spaxels), and LINER (orange spaxels). Bottom left: Spatial distribution of the BPT-classified spaxels (color-coded as in the upper panels). Empty squares mark the IFU coverage, gray squares those spaxels with continuum signal-to-noise ratio $S/N\ge1$. The red ``N'' indicates the number of AGN spaxels. Bottom right: SDSS composite image. The pink hexagon shows the MaNGA field of view.}
    \label{fig:BPT_example}
\end{figure}

Lastly, only sources with at least 20 AGN (AGN + AGN-WHAN + SF-AGN) spaxels in total were selected, since a clear separation of AGN from LINER and composite galaxies was observed at this value. This resulted in 2292 galaxies with strong indication of AGN ionization, of which there are 634 robust AGN (AGN + AGN-WHAN), 1164 SF-AGN, 114 composite, and 165 LINERs. Figure~\ref{fig:BPT_example} shows an example of the spatially resolved spaxel-by-spaxel emission line diagnostic classification.

For the purposes of this work, we have reduced the classification into only two categories: AGN, which includes robust AGN, SF-AGN, and composite, adding to a total of 1899 galaxies; and non-AGN, composed of the rest of the sample, which add to a total of 1381.\footnote{Both numbers are obtained after applying the $z\ge0.005$ and $z\le0.06$ cuts and removing the 11 erroneous galaxies.} We have considered LINERs as non-AGN since their observed emission has been shown to be not necessarily related to AGN \citep[e.g.,][]{Singh2013, Belfiore2016}. Moreover, we found that our results remain qualitatively the same if we count LINERs as AGN, although with slightly higher AGN fractions in general. With this consideration, we measured the AGN fraction as
\begin{equation}
	f_{\text{AGN}} = \frac{N_{\text{AGN}}}{N_{\text{AGN}}+N_{\text{non-AGN}}},
	\label{eq:AGN_frac}
\end{equation}
where $N_{\text{AGN}}$ and $N_{\text{non-AGN}}$ are the number of AGN and non-AGN galaxies, respectively.

\subsection{Merger classification} 
\begin{figure}
    \includegraphics[width=\columnwidth]{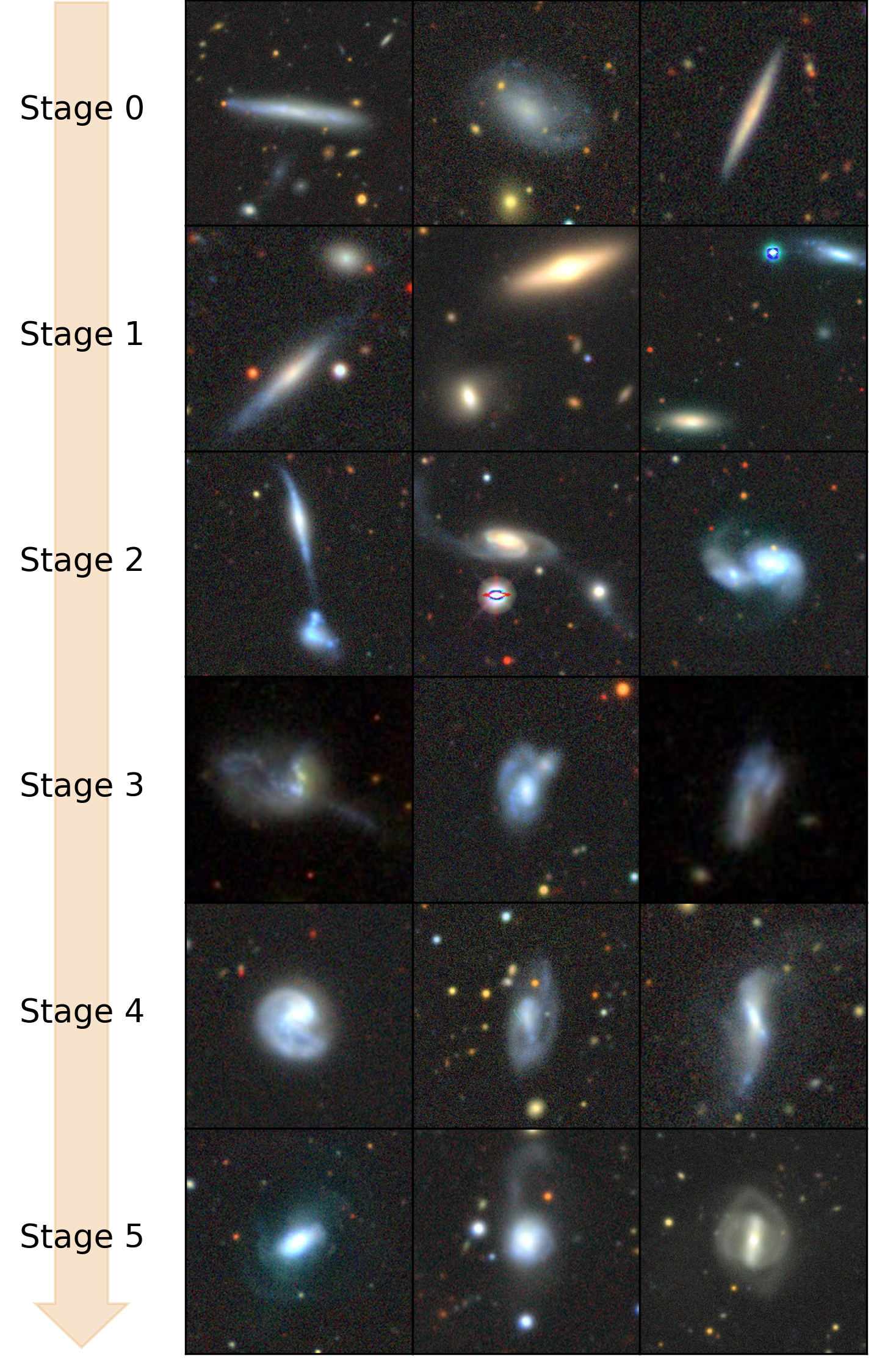}
    \caption{Example images of galaxies from our sample at different stages of the merger sequence. All images are extracted from the DESI Legacy Survey except for the first and last examples of stage 3, which are from SDSS, as this makes it easier to distinguish the nuclei for these two cases.}
	\label{fig:merger_sequence}
\end{figure}

In order to study different properties of galaxies throughout the merging process, several authors \citep[e.g.,][]{Larson2016, Smith2018, Pan2019} have defined a set of merger stages analogous to the Toomre sequence \citep{Toomre1977} and that are based mainly on visual and morphological features. In a similar effort, we have defined a merger stage sequence as follows:

\begin{itemize}
    \item Stage 0 (single galaxy): A single, non-interacting and non-paired galaxy.
    \item Stage 1 (close pair): Separated galaxies with no signs of interaction and with a projected separation of $r_{\text{p}}\leq50$ kpc and velocity separation of $|\Delta v|\leq~300~\text{km~s}^{-1}$.
    \item Stage 2 (early merger): Presumably post-first pass. Distinct galaxies with strong bridges, tails, and/or other tidal features.
    \item Stage 3 (merger): Common envelope, two distinguishable nuclei, and strong tidal features.
    \item Stage 4 (late merger): Single nucleus, highly disturbed morphology, and strong tidal features.
    \item Stage 5 (remnant): Single nucleus, less disturbed central morphology, tidal features (especially shells), and possibly off-center nucleus.
\end{itemize}
Figure~\ref{fig:merger_sequence} shows three examples of the merger sequence and additional examples of each stage are displayed in Fig.~\ref{fig:stages_extended}. We classified our sample in accordance with this sequence in two complementary steps: visual inspection and pair selection.

\subsubsection{Visual inspection} 
\label{sec:visual_inspection}
In this step we visually classify the parent sample of 3280 dwarf galaxies according to the criteria shown above, using the best image available in terms of depth and resolution. This includes images from the DESI Legacy Imaging Surveys \citep[$\text{r}_{\text{lim}}=23.54, \theta_{\text{r}}=1\overset{\prime\prime}{.}2$;][]{Dey2019}, the Hyper Suprime-Cam Subaru Strategic Program DR2 \citep[HSC; $\text{r}_{\text{lim}}=26.20, \theta_{\text{r}}=0\overset{\prime\prime}{.}8$;][]{Aihara2019}, the Hyper Suprime-Cam Legacy Archive \citep[HSCLA; $\text{r}_{\text{lim}}=24-27, \theta_{\text{r}}= 0\overset{\prime\prime}{.}4-1\overset{\prime\prime}{.}5$;][]{Tanaka2021}, SDSS \citep[$\text{r}_{\text{lim}}=22.70, \theta_{\text{r}}=1\overset{\prime\prime}{.}32$;][]{York2000}\footnote{For $\theta_{\text{r}}$ in SDSS see \url{https://www.sdss4.org/dr17/imaging/other_info/\#SeeingandSkyBrightness/}} and the Hubble Space Telescope (HST\footnote{In the case of the HST images, these may belong to different surveys or individual expositions, so there is not a defined magnitude limit value, but for reference the HST observations for the COSMOS Survey had an AB magnitude limit of 27.2 for the filter F814W and PSF width of 0$\overset{\prime\prime}{.}$1 \citep{Koekemoer2007}.}); as available in the ALADIN interactive sky atlas \citep{Bonnarel2000}, where $\text{r}_{\text{lim}}$ is the r-band magnitude depth (median $5\sigma$ detection limit in AB magnitude for a point source) and $\theta_{\text{r}}$ represents the FWHM of the image PSF in the case of DESI and SDSS, and the seeing for HSC and HSCLA.

\begin{figure}
    \includegraphics[width=\columnwidth]{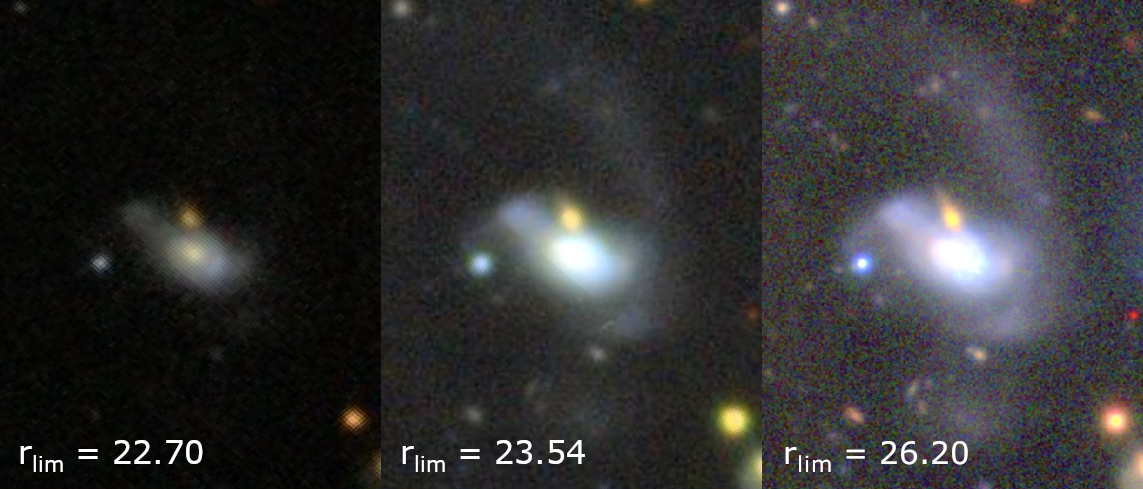}
    \caption{MaNGA galaxy 11836-3701 in composite color images from SDSS (left), DESI-Legacy (middle), and HSC-Wide (right). The bands composing the images are \textit{gri} for SDSS and HSC-Wide, and \textit{grz} for DESI-Legacy, and $\text{r}_{\text{lim}}$ indicates the AB r-band magnitude depth for each image. Tidal features that were not visible in SDSS become noticeable in DESI-Legacy and are further visible in HSC-Wide.}
    \label{fig:SDSSvDESIvHSC}
\end{figure}

With some exceptions, galaxies in stage 1 were not classified as such with the visual inspection, but rather by the pair selection procedure (see Sect.~\ref{sec:pair_selection} below). Stages 2 through 5 were classified via visual inspection. An issue to consider with this method is that the distinction between stages 1-2 and 4-5 may be held to certain ambiguity, mainly due to two reasons: the degree of disturbance and strength of the tidal features used to divide these stages depend on the criteria of the inspector, which can produce a less objective classification, and the deepest imaging survey available is not the same for all galaxies, meaning that some tidal features may be present but not detected for galaxies observed in less sensitive imaging surveys, as illustrated in Fig.~\ref{fig:SDSSvDESIvHSC}. To account for both issues we made a finer classification procedure which is detailed in Appendix~\ref{sec:Fine_stages}.

Another aspect to take into account for stages 4 and 5 is the possibility that the observed morphological disturbances are not caused by a merger process, but by tidal perturbations from nearby galaxies. In fact, simulations suggest that mergers only account for less than 20\% of morphological disturbances seen in galaxies with $M_{\star}<10^{9}~\text{M}_{\odot}$, although mergers become dominant at stellar masses $M_{\star}>10^{9.5}~\text{M}_{\odot}$ \citep{Martin2021}. Considering the stellar mass distribution of our sample (top of Fig.~\ref{fig:mass_z_dist}), we could expect that most of the observed disturbances for stages 4 and 5 are due to mergers. To test this we searched for nearby galaxies around all stage 4 and 5 mergers within a 50 kpc radius, velocity separation $|\Delta v|\leq500~\text{km}~\text{s}^{-1}$ and mass ratio $\mu\le10$, since interaction induced disturbances are not expected beyond these limits \citep{Casteels2014}. We found that only 5 out of 99 galaxies had a close companion, supporting the merger scenario. We further visually inspected these galaxies and determined that two of them presented disturbances that were more consistent with an advanced merger scenario, while the other three were more ambiguous and consequently were removed from the sample.

\subsubsection{Galaxy pair selection} 
\label{sec:pair_selection}
Given that the redshift distribution of our sample lies mainly between 0.01 and 0.06, deviations from the Hubble flow (i.e., peculiar velocities) can have a non-negligible effect on the projected distance calculations for the smaller redshift values. Therefore, we account for this prior to the search of close companions by applying the flow model of \citet{Tonry2000} to derive the redshifts corrected for peculiar velocities and linearly tapering to the observed values between $z = 0.02$ and 0.03. A similar approach can be found in \citet{Baldry2012}. Unless stated differently, from now we always use and refer to the corrected redshifts. Next, to select close pairs (i.e., stage 1 of the merger sequence) we made our main query in the NSA catalog and complemented it with SDSS DR18 only considering galaxies that have not already been classified as any other merger stage. We looked for the closest companion within a projected radius of $r_{\text{p}}=50 \text{ kpc}$ and velocity separation $|\Delta v|\leq 300~\text{km}~\text{s}^{-1}$ with respect to the MaNGA galaxies. We calculated $r_{p}$ for each galaxy considering the peculiar velocity corrected redshifts, while for $\Delta v$ we used the observed (non-corrected) redshifts. 

The limits chosen for $r_{p}$ and $\Delta v$ should include all pairs with at least a 30\% probability of ending up merging \citep{Ventou2019}. Of course, this means that a portion of close pairs might actually be flybys (i.e., will not merge). We have estimated this portion to be between 13 - 52\% of the final selection of close pairs (see Appendix~\ref{sec:pairs_to_mergers} for details). Nevertheless, the gravitational interaction effects of these flybys should not be too different from those of pairs that will end up merging. In fact, if we remove the 52\% of close pairs that are least likely to merge, our results remain qualitatively similar.

Since we are focusing on the effects of interaction on our main sample, which is composed of MaNGA galaxies, the mass ratio is computed as $\mu = M_{\star \text{M}}/M_{\star \text{c}}$, where $M_{\star \text{M}}$ and $M_{\star \text{c}}$ are the stellar masses of the MaNGA and the companion galaxies, respectively. In order to avoid extremely minor interactions, we restricted the search to companions that yielded a mass ratio $\mu\le10$ based on their NSA Petrosian stellar mass (in the case of SDSS we applied this restriction after deriving the stellar mass, as explained in Sect.~\ref{sec:mass_ratio}). We also used the mass ratio to classify as major mergers those with $\mu\leq4$, while those with $\mu>4$ were classified as minor mergers \citep[e.g.,][]{Lotz2011, Ellison2013}. When applicable, we also discarded those cases where $\chi_{r}^{2}>5$ for either the MaNGA galaxy or the companion, where $\chi_{r}^{2}$ is the goodness of fit reduced $\chi^{2}$ value resulting from the spectral energy distribution (SED) fitting (see Sect.~\ref{sec:mass_ratio}). 

In order to further increase the classification quality, the projected separation, velocity separation, mass ratio and $\chi^{2}$ restrictions were also applied to stage 2 galaxies. However since the tidal interaction can have an important effect on the measured mass ratio and luminosities, those with $\mu > 10$ or $\chi^{2} > 5$ were not reclassified as stage 0, but instead they were simply excluded from the analysis. In addition, since we are interested in dwarf-dwarf interactions, we have excluded all galaxies with non-dwarf companions (i.e., companions with $M_{\star}\geq10^{10}~\text{M}_{\odot}$) from the galaxy interaction analysis (Sects.~\ref{sec:AGN_stages} and~\ref{sec:Int_Fraction_AGN}). 

Some of the companions in stages 1 and 2 were not found in NSA nor SDSS, but were identified via visual inspection and confirmed as being interacting with spectroscopic redshifts from the NASA Extragalactic Database (NED\footnote{\url{https://ned.ipac.caltech.edu/}}) or, in those cases where the companion lies within the MaNGA IFU, from the spectra of the spaxels corresponding to the companion. We refer to these as the ``manually found'' galaxies. Finally, for a few cases where the companion galaxy did not have an available measured spectroscopic redshift and was not within the MaNGA IFU but had unambiguous signs of interaction with the respective MaNGA galaxy, the spectroscopic redshifts of the latter were assigned to the companions.

\subsection{Correcting mass ratios} 
\label{sec:mass_ratio}
The stellar masses we used throughout this work are those provided in the MaNGA survey, which in turn are extracted from the NSA catalog. In some cases, the algorithm used in the NSA catalog under-deblends close objects (which we identified in the visual inspection process), which results in a stellar mass value that is actually a combination of both objects. In addition, some of the identified companions are not present in the NSA catalog and therefore do not have an available NSA stellar mass value. 
To account for both of these problems, we took a similar approach to \citet{Steffen2023}. We derived the stellar masses of all the affected galaxy pairs fitting their ultraviolet (UV) to mid-infrared (MIR) SED with \textsc{CIGALE} \citep{Boquien2019}. In particular we use GALEX far-UV and near-UV \citep{Martin2005}, SDSS \textit{ugriz} \citep{York2000}, WISE W1-4 \citep{Wright2010} and 2MASS JHK\textsubscript{s} \citep{Skrutskie2006}.
The photometric data gathering and treatment is described in Appendix~\ref{App_Photometry}. Then we computed the mass ratio, $\mu$, of each affected galaxy pair as
\begin{equation}
	\mu = \frac{M_{\star\text{M}}^{\text{SED}}}{M_{\star\text{c}}^{\text{SED}}},
	\label{eq:m_ratio}
\end{equation}
where $M_{\star\text{M}}^{\text{SED}}$ and $M_{\star\text{c}}^{\text{SED}}$ are the stellar masses obtained from the SED fitting of the MaNGA and companion galaxies, respectively. We then combined the mass ratio with the NSA stellar masses, $M_{\star\text{NSA}}$, to compute the individual stellar masses of the under-deblended close pairs: 
\begin{equation}
	M_{\star\text{M}}=M_{\star\text{NSA}}\frac{\mu}{1+\mu},
	\label{eq:mass1}
\end{equation}
\begin{equation}
	M_{\star\text{c}}=M_{\star\text{NSA}}\frac{1}{1+\mu}.
	\label{eq:mass2}
\end{equation}
The following is for the case of the companions with missing NSA stellar masses:
\begin{equation}
	M_{\star\text{c}}=\frac{M_{\star\text{NSA}}}{\mu}
	\label{eq:mass4},
\end{equation}
while the MaNGA stellar masses remained $M_{\star\text{M}}=M_{\star\text{NSA}}$.

\subsection{The galactic environment} 
In order to further explore the impact of the environment on AGN triggering and disentangle it from the effects of close interactions/mergers, we need to characterize the environment. There are several approaches in the literature to this end \citep[e.g.,][etc.]{Cooper2005, Baldry2006, Geha2012, Pearson2016, Yang2007, ArgudoFernandez2015}. To have a wider perspective and be able to check for consistency, we have selected three different environmental factors for a galaxy: being isolated, in a void, or part of a group or cluster.

\subsubsection{Isolated galaxies} 
The presence of a nearby massive galaxy has been used as proxy to study the impact of the environment and the star formation rate (SFR) of dwarf galaxies as isolated and non-isolated \citep[e.g.,][]{Geha2012, Stierwalt2015, KadoFong2019}. In particular, \citet{Geha2012} found that the fraction of quenched dwarf galaxies strongly increases with lower projected separations to a massive galaxy ($M_{\star}\geq2.5\times10^{10}~\text{M}_{\odot}$), and remains constant at separations higher than 1.5 Mpc. Following the parameters and results of their work, we define a dwarf isolated galaxy as not having any massive galaxies within $|\Delta v|\leq1000~\text{km~s}^{-1}$ and 1.5 Mpc of projected radius. 

The search for nearby massive galaxies was first performed within the NSA catalog considering the Petrosian stellar mass and then in the SDSS database using the CasJobs tool\footnote{\url{https://skyserver.sdss.org/casjobs/}} considering the \texttt{logMass\_noMassLoss} column from the \texttt{stellarMassStarformingPort} table for the stellar mass value. A few potential massive neighbors found in SDSS had no available stellar mass values. For these cases we estimated the stellar mass using the formula and parameters provided in \citet{Roediger2015} for \textit{g-r}. 

Since there might be a few massive neighbors that lie outside the footprint of SDSS (and therefore of NSA) we also searched in the 2MASS Extended Source Catalog \citep[XSC;][]{Jarrett2000} for sources with absolute magnitudes $M_{Ks}\le 22.86$, which corresponds to the massive galaxy threshold assuming $(M_{\star}/L)_{K_{S}}=0.95~\text{M}_{\odot}/\text{L}_{\odot}$ and $M_{K_{S}\odot}=3.32$ \citep{Bell2003}. The redshifts for the 2MASS sources were obtained by crossmatching with several spectroscopic surveys\footnote{Along with SDSS, these surveys are the sources for the redshifts in NSA.}: NED, the 100\% complete Arecibo Legacy Fast ALFA Survey \citep[ALFALFA;][]{Haynes2018}, the 2dF Galaxy Redshift Survey \citep[2dFGRS;][]{Colless2001,Colless2003}, the 6dF Galaxy Redshift Survey \citep[6dFGS;][]{Jones2004,Jones2009} and the CfA Redshift Survey \citep[ZCAT\footnote{\url{https://www.cfa.harvard.edu/~dfabricant/huchra/zcat/}};][]{Huchra1995}. The applied k-correction was obtained using the code provided by \citet{Chilingarian2012}, and the extinction was derived using the color excess $E(B-V)$ values of \citet{Schlafly2011}, provided by the IRSA Galactic Dust Reddening and Extinction service\footnote{\url{https://irsa.ipac.caltech.edu/applications/DUST/}} and using $R_{K_{S}}=0.306$ \citep{Yuan2013}. 

Finally, we matched the results from the 2MASS XSC with the NSA and SDSS and removed those galaxies that had been previously rejected in the NSA and SDSS. This yields a total of 2329 ($\sim71\%$ of the sample) non-isolated galaxies.

\subsubsection{Galaxies in voids} 
Whether galaxies are located (or not) within a void has been observed to have an impact on AGN activity \citep[e.g.,][]{Constantin2008, Mishra2021, Ceccarelli2022}, which should be taken into account when measuring the AGN fraction.

It is important to highlight the difference between isolated and void galaxies: voids are generally defined based on the density and extension of a region, while isolation depends simply on the presence of a galaxy with stellar mass $M_{\star}\geq2.5\times10^{10}~\text{M}_{\odot}$ within defined projected radius and radial velocity limits. This means that an isolated galaxy will not necessarily be in a void. Conversely, a galaxy may be within a void region, but not isolated. Hence, these two criteria are somewhat similar but nevertheless complementary.

To classify our galaxies as being located within a void (or not) we crossmatched our galaxy sample with that of \citet{Pan2012}, who constructed a catalog of galaxies in voids using the VoidFinder algorithm \citep{Hoyle2002}. All the matches were marked as being in a void in our sample. Next, to distinguish those galaxies that are not in a void from those that have non valid data, we crossmatched the rest of the galaxies from our sample to the parent catalog of the void sample, which is the Korea Institute for Advanced Study Value-Added Galaxy Catalog (KIAS-VAGC). The positive matches were then marked as not being in a void, while the rest of the galaxies were marked as non-valid (and will hence be excluded from any analysis regarding voids). This results in a total of 1013 ($\sim31\%$) galaxies in voids, 2157 ($\sim66\%$) not in voids and 110 ($\sim3\%$) non-valid.

\subsubsection{Groups and clusters} 
It is thought that being part of a group or cluster also has an effect on the AGN fraction of dwarf galaxies \citep[e.g.,][]{CalderonCastillo2024}, although there are mixed results on this \citep[e.g.,][]{Siudek2022}.

Similarly to the voids, we find out if each of our galaxies belongs to a group or a cluster by crossmatching our sample with the Self-calibrated Galaxy Group Catalog \citep[hereafter the Group Catalog;][]{Tinker2021, Tinker2022}, which includes the group richness (i.e., the number of members in the group) for each galaxy, which we denote as $G$. Next, we divided the results in three categories: Field galaxies ($G=1$, meaning the galaxy is not part of a group), Group galaxies ($2\leq G \leq 50$) and Cluster galaxies ($G > 50$). We chose 50 as it is a typical value for the transition from groups to clusters \citep{Abell1958}. This resulted in 1937 ($\sim59\%$) field galaxies, 652 ($\sim20\%$) group galaxies and 369 ($\sim11\%$) cluster galaxies. The 322 ($\sim10\%$) galaxies that had no match in the Group Catalog were excluded from any analysis regarding group richness.

\subsection{Sources of incompleteness and biases} 
Given the parent surveys and magnitude ranges used in this work, there are several sources of incompleteness that can potentially affect the quality and relevance of our results, especially when considering the method for classifying galaxies in stage 1 (close pairs). Here we discuss how we addressed each of them.

\subsubsection{Magnitude limits and overall spectroscopic completeness} 
\label{sec:mag_spec_completeness}
\begin{figure}
    \includegraphics[width=\columnwidth]{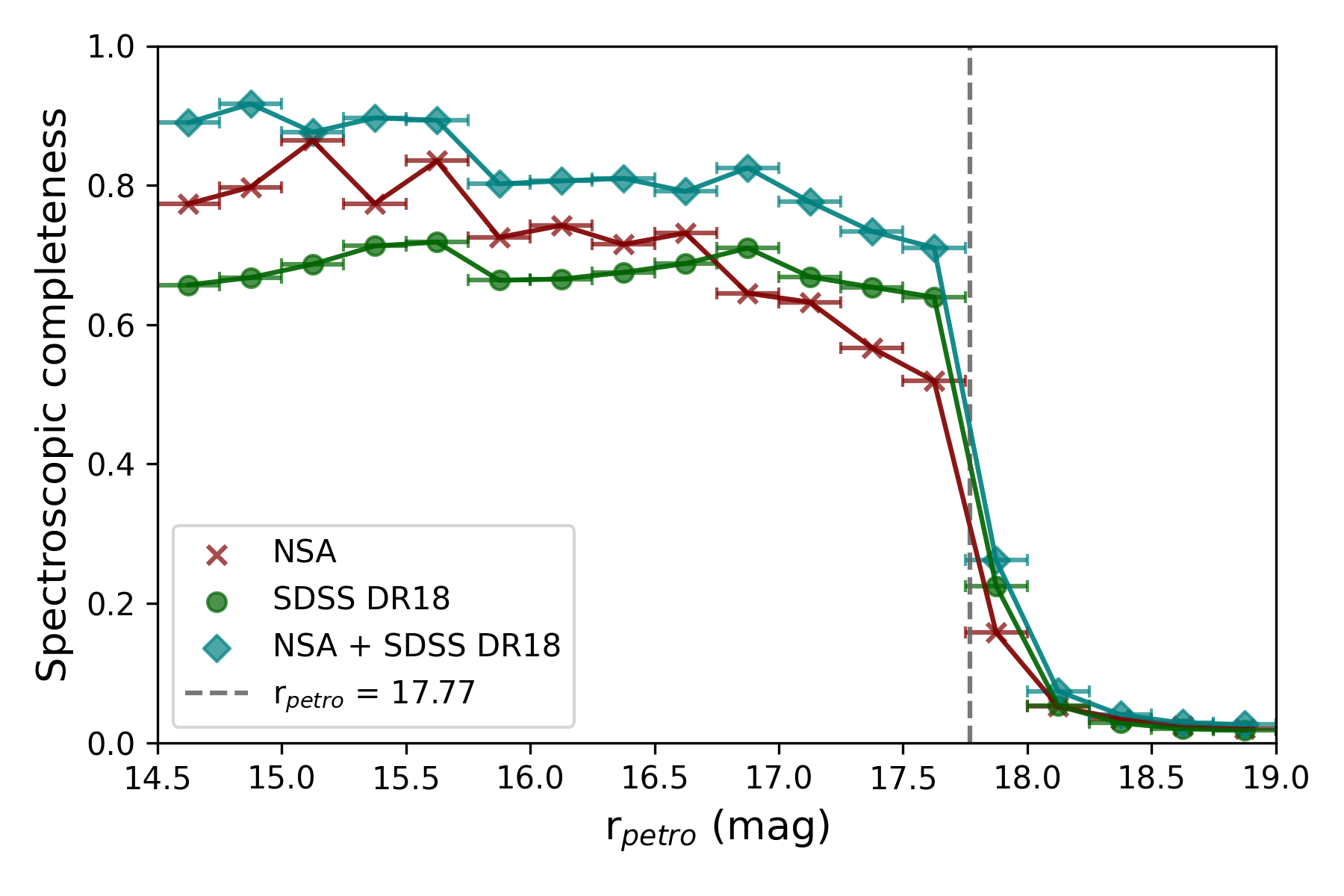}
    \caption{Spectroscopic completeness vs. r-band Petrosian magnitude in SDSS DR18 (green dots) and NSA (red crosses) and their combination (blue diamonds). The vertical line at $\text{r}_\textit{petro} = 17.77$ shows the magnitude selection limit for the SDSS spectroscopic sample. The spectroscopic completeness was computed as the number of galaxies with spectra in SDSS or NSA respectively, over the number of galaxies with photometry in SDSS DR18 within a 300\arcsec radius of each galaxy of our sample.}
    \label{fig:mag_completeness}
\end{figure}

The majority of the companion galaxies we found come from NSA and/or SDSS, the latter being the main (but not the only) source of redshift values of the former. Consequently, there is a very significant drop in spectroscopic completeness (computed as $f_{\text{s}}=N_{\text{z}}/N_{\text{p}}$, where $N_{\text{z}}$ and $N_{\text{p}}$ are the number of spectroscopic and photometric\footnote{Since the photometry of NSA is based on SDSS, the number of photometric objects is the same for both surveys.} galaxies, respectively.) for r-band magnitudes $r \gtrsim 17.77$, which is the magnitude limit for the SDSS spectroscopic sample \citep{Strauss2002}. This is illustrated in Fig.~\ref{fig:mag_completeness}, which shows the spectroscopic completeness vs. the magnitude for both SDSS and NSA within a 300\arcsec radius\footnote{We selected this radius since it provides sufficient sources to have a reliable measure of completeness while being very specific for our sample.} of each galaxy of our sample. This magnitude limit results in an underrepresentation of galaxies with lower brightness (a.k.a. the Malmquist bias), and consequently of companions with lower masses and pairs of higher mass ratios (i.e., minor mergers), especially at higher redshifts. Table~\ref{tab:missing_comp_tab} shows estimations of the number of undetected close companions due to flux limits for $\mu \leq 4$, $4<\mu \leq 10$ and the total. The derivation of these values is mainly based on the magnitude difference and mass ratios of galaxy pairs, and is detailed in Appendix~\ref{missing_comp_estimation}. 

\begin{figure}
    \includegraphics[width=\columnwidth]{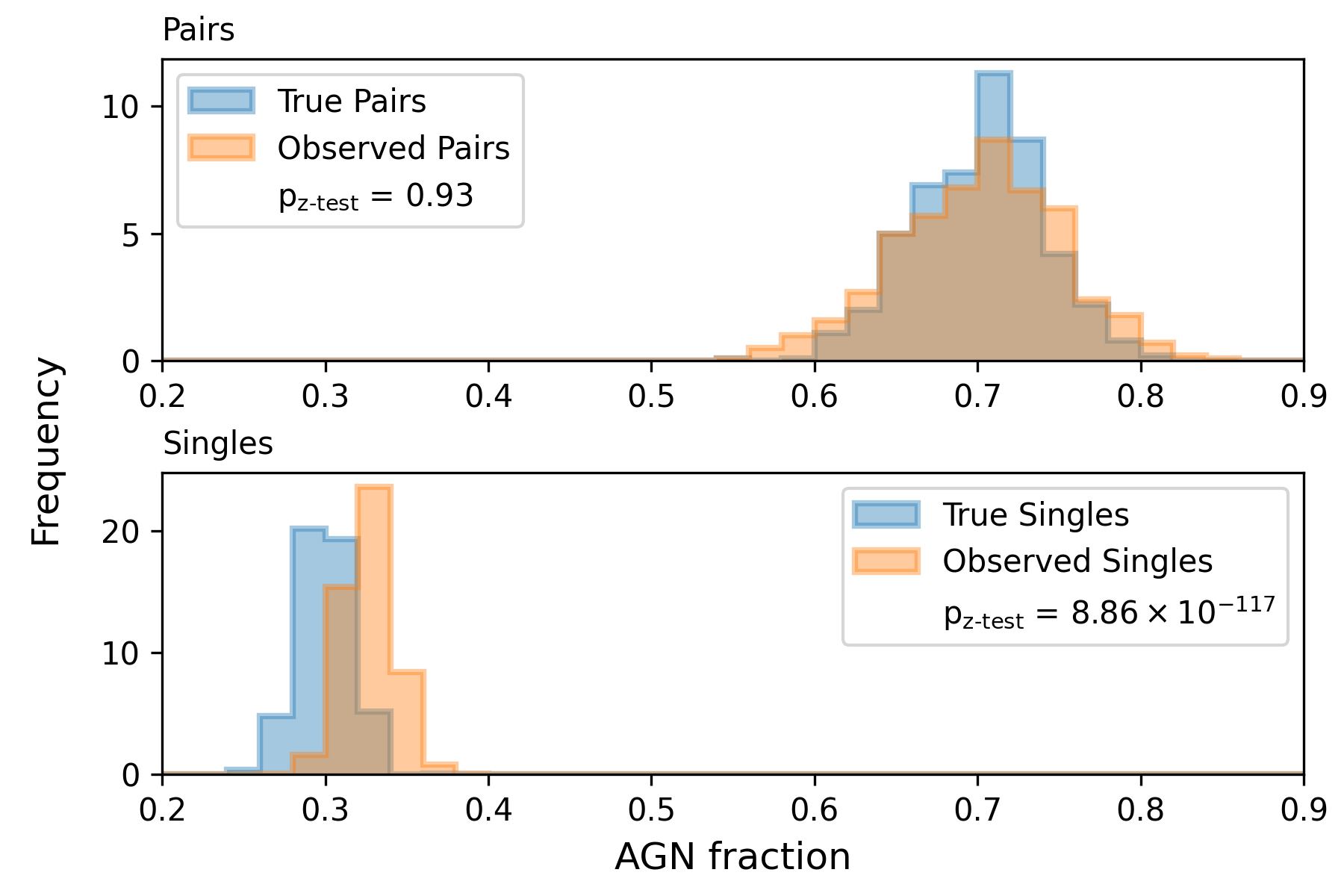}
    \caption{Simulated AGN fraction distribution of true and observed galaxy pairs and single galaxies when assuming the probabilities stated in the text for a sample size of 1000 galaxies and over 500 iterations. The term $\text{p}_{\text{z-test}}$ indicates the p-value obtained from a two sample z-test performed on the true and observed distributions to test if they have the same mean value.}
    \label{fig:pair_simulation}
\end{figure}

In principle, this underrepresentation could be compensated for by applying weights (analogous to the approach explained in Sect.~\ref{sec:Fiber_collisions}). However, only a small portion ($\sim8\%$, 22 galaxies in total) of the detected companions have magnitudes $r > 17.77$. Compensating weights based on such a small subset could lead to a more inaccurate representation of the underlying population. Moreover, we expect that this bias has little to no effect in our results, considering the following argument: If the magnitude limit affects both the AGNs and non-AGNs of our sample equally, we should miss companions of AGN and non-AGN galaxies proportionally to the AGN fraction of the full sample. In other words, the AGN fraction in observed pairs should be equal to the AGN fraction in true pairs. This argument remains valid whether there is a correlation between close pairs and being an AGN or not. However, under the assumption that there is, in fact, such a correlation, the AGN fraction of observed single galaxies could differ from that of true single galaxies. This is because when counting as single galaxies those that are observed as single, but are in fact pairs, the included AGNs and non-AGNs are not proportional to the AGN fraction of the full sample. This situation is illustrated in Fig.~\ref{fig:pair_simulation}, which shows a simulation of the AGN fraction distribution for true and observed galaxy pairs and non-pairs assuming $\text{p}_{\text{TP}} = 0.15$, $\text{p}_{\text{obs}} = 0.6$, and $\text{p}_{\text{P-AGN}} = 0.7$, where $\text{p}_{\text{TP}}$ is the probability of having a true pair, $\text{p}_{\text{obs}}$ is the probability of observing a pair and $\text{p}_{\text{P-AGN}}$ is the probability that a paired galaxy is an AGN (assuming a positive correlation). Based on these probabilities we generate a sample of 1000 galaxies to measure the AGN fraction, and then repeat this process 500 times to obtain the AGN fraction distribution. We note that the p-value $\text{p}_{\text{z-test}}$ obtained from the z-test performed to compare the means of the true and observed distributions is well above the commonly accepted significance level $\alpha=0.05$, suggesting the true and observed pairs distributions have the same mean, although the observed distribution has a higher uncertainty. The behavior remains qualitatively the same for different values of $\text{p}_{\text{TP}}$, $\text{p}_{\text{obs}}$ and $\text{p}_{\text{P-AGN}}$ (even when allowing them to vary for each iteration), with the only relevant difference being the separation for the observed and true singles and the dispersion of observed pairs. In fact, the values used for this plot were chosen to accentuate this difference, since the values we have estimated yield an even smaller (but still statistically significant) difference. Fortunately, not only does this difference tend to be small, but also, since it is present only in the AGN fraction of single galaxies, it is averaged out when combining with the other control galaxies (see Sect.~\ref{control_sample}). 

Another thing to be taken into account is how this magnitude limit affects the early mergers/pairs fraction when comparing the AGN and non-AGN samples. Since the decrease in the early mergers/pairs fraction due to flux limits is proportional to the number of true early mergers/pairs, the stronger the AGN-merger correlation is, the more prominent is the effect of the undetected pairs. This means that any difference found between the early mergers/pairs fractions of the AGN and non-AGN samples will be slightly attenuated. Nevertheless, based on our estimations of the total of undetected companions (see Table~\ref{tab:missing_comp_tab}), we expect this effect to be small.

Lastly, we have to consider the overall completeness. We have measured a mean spectroscopic completeness of $\sim0.67$ at $r \leq 17.77$ for both SDSS and NSA separately, and $\sim0.8$ for their combination. This means that even below the magnitude limit, there is a 20\% chance that we miss a close companion and consequently underestimate the distance to the closest companion or classify a close pair (i.e., stage 1) as a single galaxy (i.e., stage 0). Although this value is not negligible, it still represents a minority. More importantly, based on the argument above, it should not systematically affect our results for the AGN fraction.

\begin{table}
    \centering
        \caption{Estimated numbers of undetected close companions.}
    \label{tab:missing_comp_tab}
    \begin{tabular}{lcc} 
        \hline
          Mass ratio range & Observed & Undetected \\
        \hline
        $\mu \leq 4$ & 218 (0.95) & 12 (0.05) \\
        $4<\mu \leq 10$  & 36 (0.65) & 19 (0.35) \\
        Total ($\mu \leq 10$) & 254 (0.89) & 31 (0.11) \\
        \hline
    \end{tabular}
    \tablefoot{The rows correspond to the mass ratio ranges of major ($\mu \leq 4$) and minor ($4<\mu \leq 10$) mergers and the total. The number inside the brackets is the proportion corresponding to the total of the row.}
\end{table}

\subsubsection{Red galaxies and the low-mass end} 
\label{sec:low_mass_end}
A consequence of the selection method for the MaNGA targets is that it results in an underrepresentation of red galaxies for stellar masses $M_{\star}<10^{10}~\text{M}_\odot$, especially at the lowest mass end \citep{Wake2017}. Although this issue is somewhat mitigated in the Primary+ sample of MaNGA (see Sect.~\ref{sec:data_class}), the incompleteness may still cause a severe decrease in the red dwarf galaxy population at the low-mass end. According to \citet{Wake2017}, the sample is virtually complete for stellar masses $M_{\star}^{\text{p+}}>1.2 \times 10^{9}~\text{M}_{\odot}$ and $M_{\star}^{\text{s}}>3.7 \times 10^{9}~\text{M}_{\odot}$\footnote{These values have already been adjusted to our cosmology.} for the MaNGA Primary+ and Secondary samples, respectively. Consequently, we remove all galaxies with stellar masses below these limits from our main sample. This affects 24 galaxies.

After applying these cuts, our sample is still likely dominated by blue galaxies, as suggested by the stellar mass distributions of red, green, and blue galaxies in the Primary+ sample presented by \citet{Wake2017}. However, we note that, as pointed out by \citet{Mezcua2024}, the median of the \textit{B-V} color of the sample used in this work is redder than that of dwarf AGN galaxies identified through other methods \citep[e.g.,][]{Reines2013, Mezcua2018, Mezcua2019b, Reines2020}.

\subsubsection{Fiber collisions (SDSS/NSA)} 
\label{sec:Fiber_collisions}
Another type of incompleteness that has a direct impact on galaxy close pairs comes from fiber collisions within the SDSS multi-object spectrograph, which permits a minimum angular separation of 55\arcsec \citep{Blanton2003}, meaning that the spectra of galaxy pairs with separations lower than this cannot be acquired simultaneously using a single spectroscopic plate. This minimum separation corresponds to 50 kpc at a redshift of $z \approx 0.048$, and hence most of our sample is affected by it. There are regions, however, in which two or more plates overlap, meaning that a fraction of galaxy pairs with separations below 55\arcsec are recovered. 

\begin{figure}
    \includegraphics[width=\columnwidth]{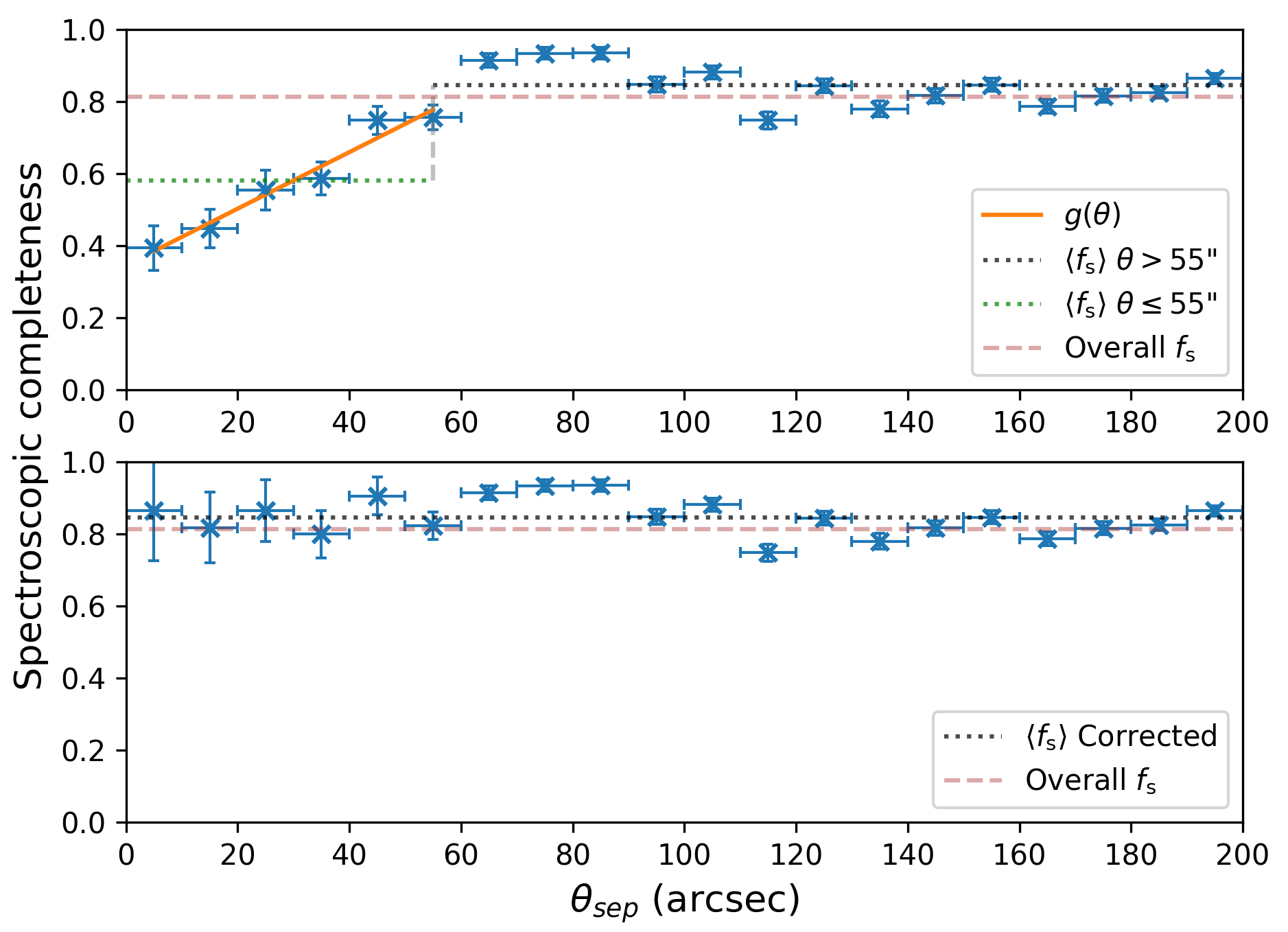}
    \caption{Spectroscopic completeness vs. angular separation for the SDSS DR18 and NSA combined. Top: Data before applying the weight correction. The spectroscopic completeness noticeably decreases from $\theta = 55^{\prime\prime}$ to smaller separations, while for $\theta > 55^{\prime\prime}$ it converges toward the overall spectroscopic completeness ($f_{s} \approx 0.83$). The solid orange line shows the fit to the data below $\theta = 55^{\prime\prime}$. To improve the fitting quality and reduce error sizes, we used 10\arcsec wide bins. Consequently, we also include a few galaxies between 55\arcsec and 60\arcsec for the fitting. Bottom: Data after applying the weight correction. The spectroscopic completeness was computed in the same way as in Fig.~\ref{fig:mag_completeness}. The error bars were computed by error propagation.}
    \label{fig:fiber_collision}
\end{figure}

To account for the remaining incompleteness we took an empirical approach similar to that of \citet{Patton2008}, but adapted to our situation. First we made a spectroscopic sample by searching for all galaxies with spectra within a radius of 300\arcsec of each galaxy of our sample in both SDSS DR18 and NSA, combining the results of both samples and removing all redundant objects. We then made a photometric sample, by searching for all galaxies with photometry in SDSS DR18 within the same region. Since NSA is restricted to $z\leq0.15$, in order to maintain consistency we removed all objects with $z>0.15$ from both the spectroscopic and photometric samples (for the photometric sample we first matched it to the spectroscopic sample to know which objects were beyond the redshift limit). After this we removed from both samples all the galaxies that are in our MaNGA sample, and added the manually found spectroscopic neighbors in order to prevent overcorrection. Next we computed the spectroscopic completeness (as explained in Sect.~\ref{sec:mag_spec_completeness}) at different angular separations, as shown in the top panel of Fig.~\ref{fig:fiber_collision}. As expected, completeness decreases below $\theta = 55^{\prime\prime}$. To compensate for this, we multiply each pair below $\theta<55^{\prime\prime}$ by a weight $w_{\theta} = f_{\text{s}}/g(\theta)$, where $g(\theta)$ is a function representing the best fitting line below $\theta = 55^{\prime\prime}$. In turn, all galaxy pairs with $\theta > 55^{\prime\prime}$ have been given a weight of $w_{\theta}=1$. We note that the bin width used in Fig.~\ref{fig:fiber_collision} is of 10\arcsec. This was done to reduce the uncertainties and improve the fitting quality. Consequently, the fitting includes some galaxies that are between 55\arcsec and 60\arcsec. Upon several tests, we found that was the best trade-off, as the fitting parameters show negligible differences. The bottom panel of Fig.~\ref{fig:fiber_collision} shows the effect of applying the correcting weights. 

\subsubsection{Close pair spaxel overcount} 
Another potential bias that we detected comes from the AGN selection method used in \citet{Mezcua2024}. Since they only selected sources with more than 20 AGN spaxels in their final classification, close pairs in which the companion is partially or completely within the MaNGA field are more likely to be included, since the companion spaxels would be counted as well. Upon visually checking all potentially affected galaxies, we found only one case (9490-12701; shown in Fig.~\ref{fig:Spaxel_overcount_example}) where the main galaxy has less than 20 AGN spaxels. However, we did not exclude it from the sample since this galaxy is particularly small with respect to its MaNGA field (possibly due to an IFU allocation based on an under-deblended close pair), meaning that the proportion of AGN spaxels is relatively high.

\subsection{Control sample}\label{control_sample} 
In order to isolate the effect of close pairs/mergers on AGN fraction and further account for possible remaining biases, we have constructed a control sample matched in stellar mass and redshift. For this purpose, we separated our sample into two subsets: the interacting galaxy sample, composed of galaxies in stages 1-5, and the control pool, composed of galaxies at stage 0. To account for potential environmental effects, we add a restriction so that galaxies can only be matched to those with the same condition of being isolated and being in a void\footnote{Here we have omitted the field, group or cluster criterion because the matching became too restricted, resulting in some galaxies with zero matches.}, then find the best simultaneous match in stellar mass and redshift to each galaxy of the interacting sample from the control pool and run a Kolmogorov–Smirnov (KS) test between the interacting sample and the matched controls. To be consistent with other works \citep[e.g.,][]{Ellison2011}, if the KS p-value is above 0.3, we keep that control sample and repeat the process to find the next best match. This process was repeated until the KS p-value reached a value of 0.3 or lower, or we reached ten iterations. In our case, the second condition was met first. We note that although an interacting galaxy can only be matched to a certain control galaxy once, the control galaxies can be matched to several interacting galaxies. 

\section{Results and discussion}
\label{sec:Results}
\subsection{AGN fraction throughout the merger process} 
\label{sec:AGN_stages}
\begin{figure}
        \includegraphics[width=\columnwidth]{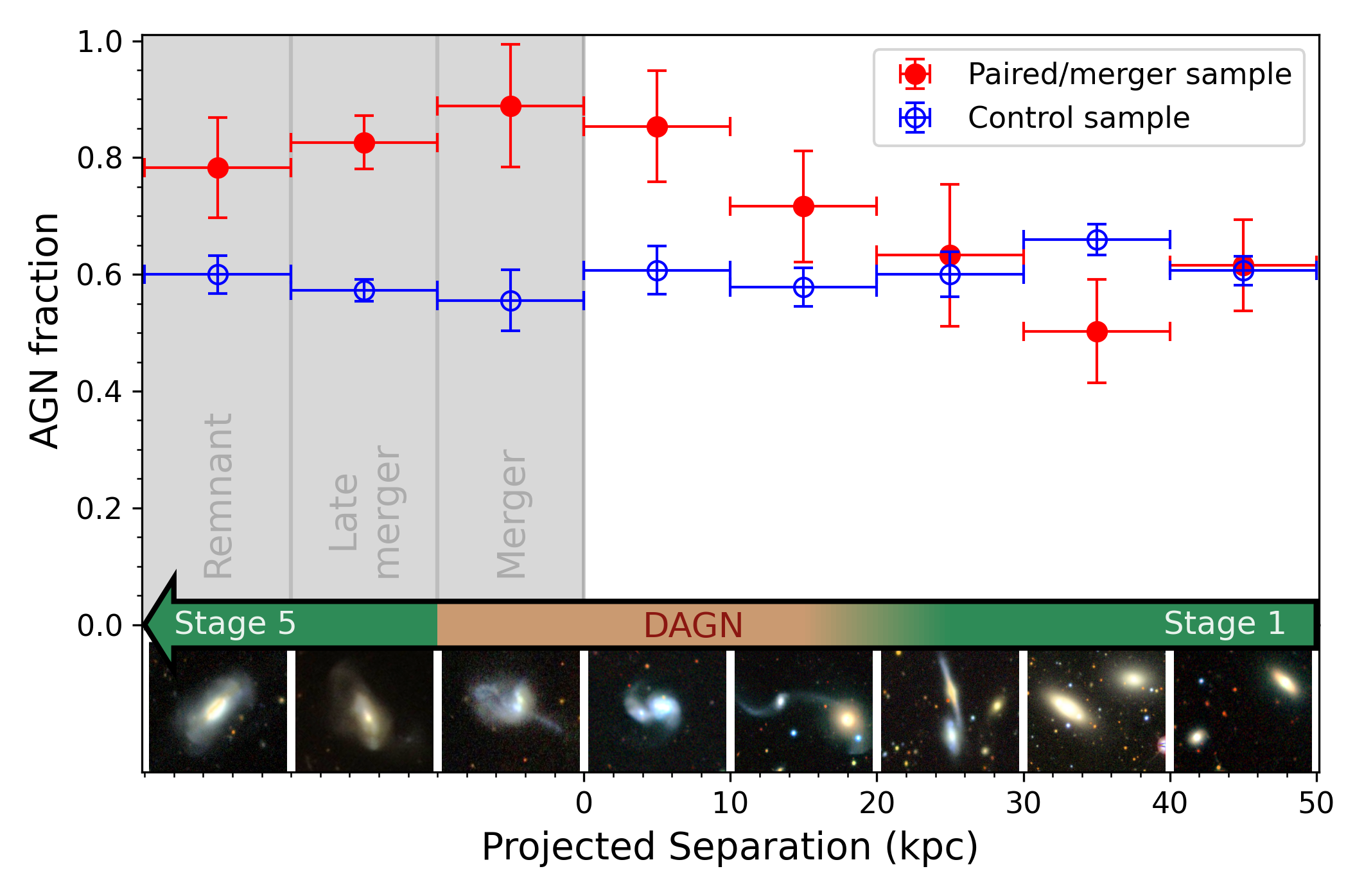}
        \label{fig:AGN_proj_sep}
        \includegraphics[width=\columnwidth]{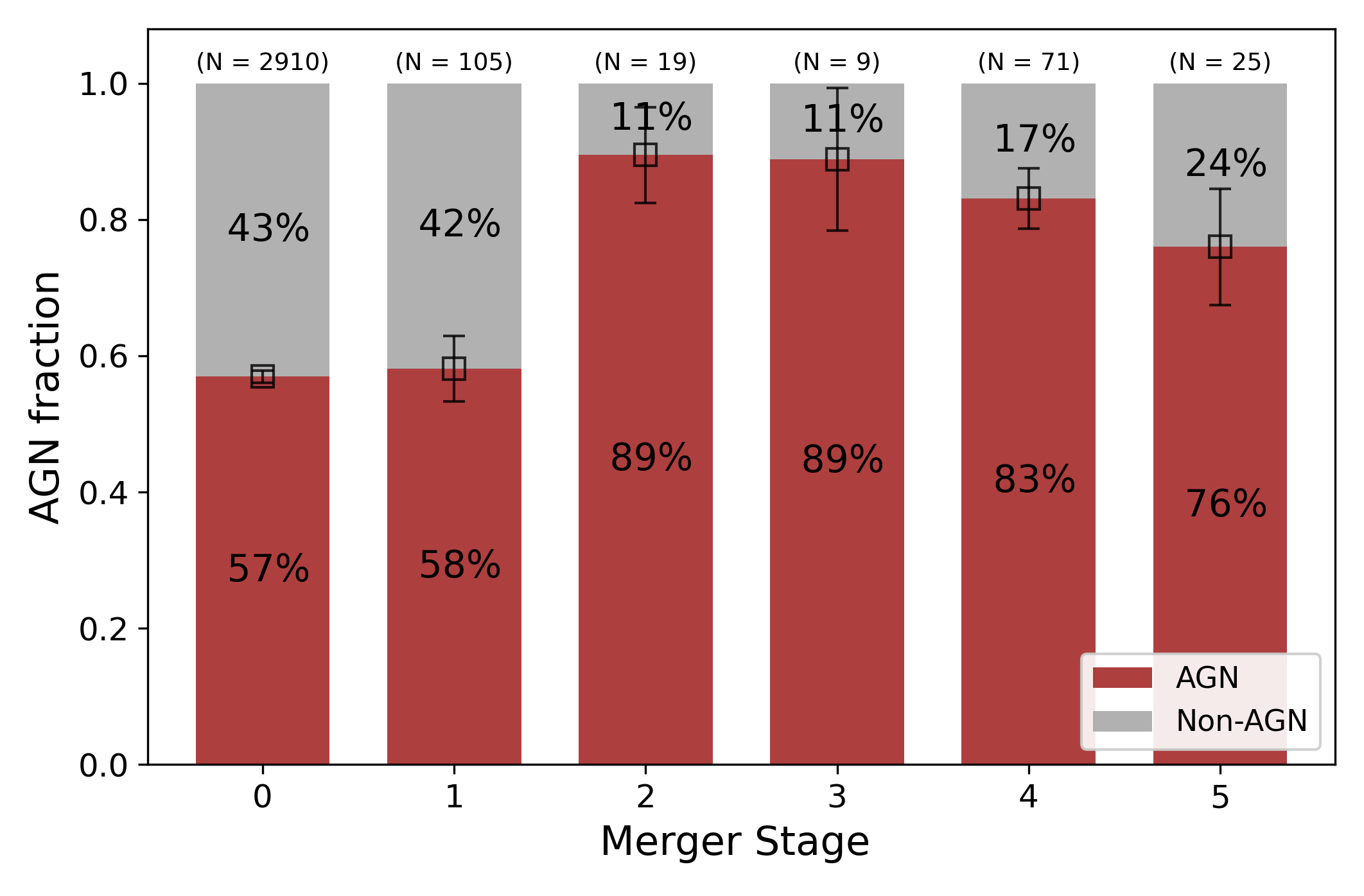}
        \label{fig:AGN_stages}
        \caption{Top panel: AGN fraction vs. the projected separation for the paired and merging galaxies sample (filled red points) and its control group (open blue points). The horizontal bars show the bin width, while the vertical bars show the binomial error of each bin. The gray bins at the left of the 0 kpc mark correspond to stages 3, 4, and 5 from right to left, respectively. The green arrow shows the direction of evolution of the merger process, and the orange region inside of it indicates the range in which we could expect to find dual AGNs (we note that the right end is a gradient, since the condition of being a dual AGN will depend on the applied definition and more factors than only the projected separation.). The galaxy images below the arrow are representative examples of the corresponding bin. We note that stages 1 and 2 are mixed in the projected separation region. Bottom panel: AGN fraction at each merger stage, shown as a percentage. The error bars show the binomial error of each bin. The numbers on top indicate how many AGNs are in the corresponding bin. We note that the AGN fraction for stages 3-5 might slightly vary from the top plot because a few galaxies were excluded from it, since they did not have a valid value for the Void criterion and hence could not be matched to control galaxies.}
        \label{fig:merger_AGNfraction}
\end{figure}

We first compare the AGN fraction of paired/merger galaxies to the respective control group at different bins of projected separation between pairs, as shown in the top panel of Fig.~\ref{fig:merger_AGNfraction}. The AGN fraction of galaxy pairs at projected separations below 20 kpc (where pairs with tidal features become the dominant population) starts to increase notoriously, reaching a value of 0.85 (1.41 times the AGN fraction of the control group in the respective bin) below 10 kpc. Then, the AGN fraction peaks with 0.89 at stage 3 (merger) and remains high throughout the rest of the merger process, although slightly decreasing toward the final stage. This behavior is consistent with that observed for massive galaxy dominated samples \citep[e.g.,][]{Ellison2011, Satyapal2014, Barrows2023}, but paired massive galaxies start to show AGN excess at higher separations. This indicates that the AGN-merger connection does extend to the low-mass domain, but the triggering separation scales down with stellar mass. Interestingly, when comparing the AGN fraction of all merger stages (bottom panel of Fig.~\ref{fig:merger_AGNfraction}) we find a distinct AGN fraction boost for stages 2-5. To confirm this we grouped stages 0-1 and 2-5 and performed a $\text{$\chi$}^{2}$-test of independence with which we found a very significant statistical difference between both groups ($p_{\mathrm{\chi}^{2}}=6\times10^{-9}$)\footnote{We also performed a post hoc test, which yielded seemingly consistent results. However, some of the differences become much less significant due to the small size of some individual categories.}. Although this is consistent with the top panel of Fig.~\ref{fig:merger_AGNfraction}, as we expect to find more galaxies in stage 2 than stage 1 at smaller separations (and vice-versa for larger separations), it also suggests that only interactions from the first pass onward trigger AGNs. \par
We further investigated the effect of potential contamination by SF emission in the AGN selection, and repeated the entire process removing the SF-AGN galaxies from our sample (Fig.~\ref{fig:merger_AGNfraction_NoSF-AGN}). After removing the SF-AGN, the AGN fractions are lower in general but a similar trend remains, having a higher AGN population for close pairs/mergers. In this case the same $\chi^{2}$-test yielded a smaller but still significant difference ($p_{\chi^{2}}=0.04$), although the AGN fraction boost seems to already take place at stage 1 (i.e., before the first pass). This is possibly due to the fact that galaxies at stage 1 have the lowest SFR values in the merger sequence (see Fig.~\ref{fig:SFR_stages}), meaning that the AGN fraction at this stage is likely less affected by the removal of SF-AGN from the sample. This could also explain the notorious drop for stage 4. 

Although the literature about AGNs in dwarf galaxy mergers is very scarce, we found that our results are in agreement with \citet{Micic2024}, who reported an enhancement in the frequency of X-ray identified AGN of grouped/paired dwarf galaxies when compared to non-interacting dwarf galaxies. Similarly, \citet{LaMarca2024} found an AGN excess in mergers compared to non-mergers based on MIR, X-ray and SED AGN identification methods, and in all their redshift bins. While their work is not constricted to dwarf galaxies, their stellar mass distribution vs. redshift plot seems to indicate that in their lowest redshift bin ($0.1\leq z<0.31$) the stellar masses of the dominant population are within $10^{9} - 10^{10} ~\text{M}_{\odot}$. This AGN excess in mergers has also been observed at much higher redshifts ($5.3\leq z<6.7$) by \citet{Duan2024} based on BPT and SED selection methods, and with a sample mostly composed of galaxies with $M_{\star}<10^{10}~\text{M}_{\odot}$. Considering the variety of sources and AGN selection methods, the evidence supporting an AGN-merger connection in the dwarf galaxy regime is quite robust, and we find that possibly this connection becomes observable from the first pass of the merger.

\subsection{The interacting fraction of AGNs} 
\label{sec:Int_Fraction_AGN}
\begin{figure}
    \includegraphics[width=\columnwidth]{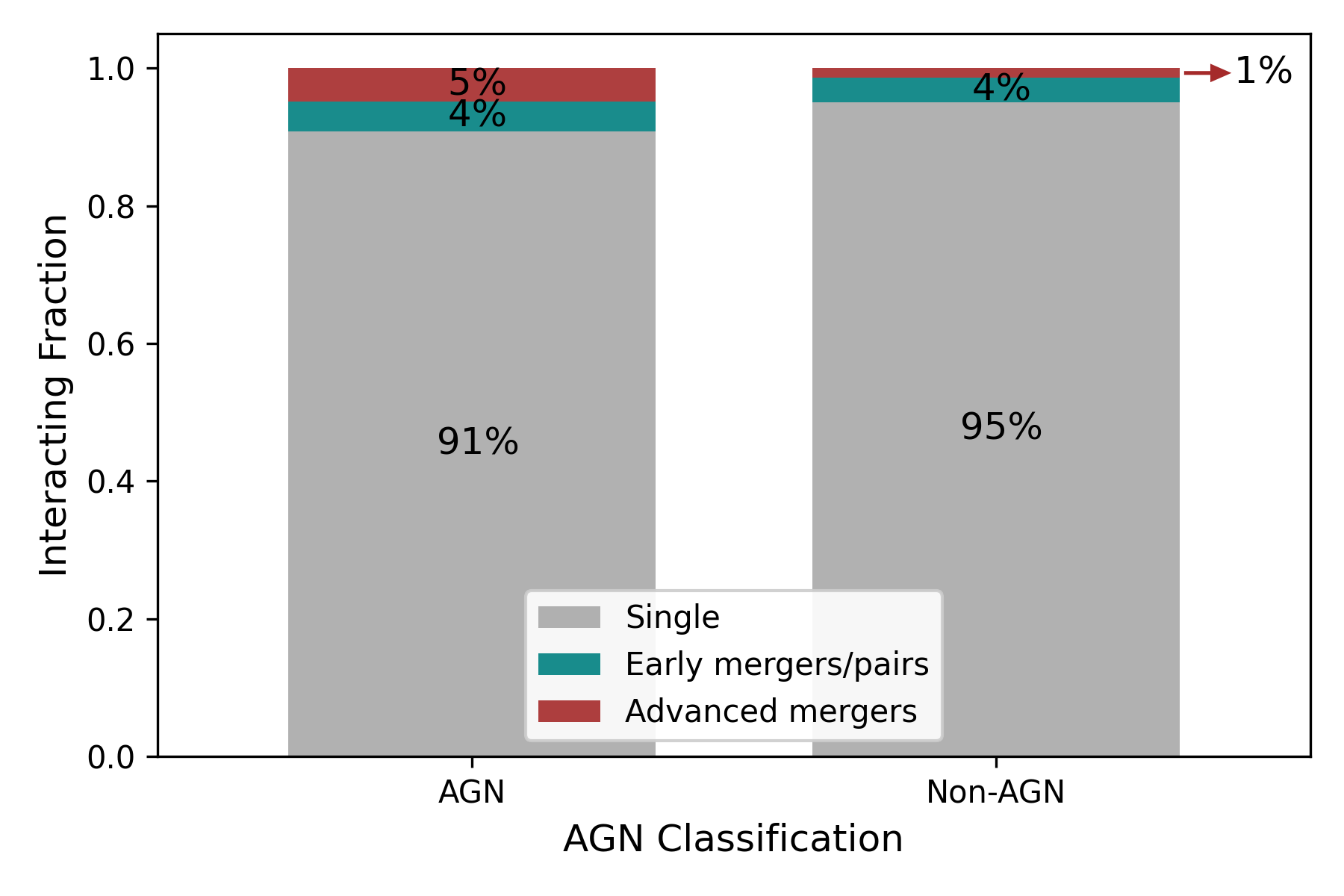}
    \caption{Fraction of single (stage 0), early mergers and pairs (stages 1-2) and advanced mergers (stages 3-5) in the AGN and non-AGN samples.}
    \label{fig:intfrac_AGNnon}
\end{figure}

In the previous section we showed that dwarf-dwarf mergers can trigger AGNs; however, this is not a complete view on the merger-AGN connection. We now combine our samples the other way around. Figure~\ref{fig:intfrac_AGNnon} shows the fractions of single galaxies (stage 0), pairs and early mergers (stages 1 and 2) and advanced mergers (stages 3-5) for the AGN and non-AGN samples, respectively.
At first glance, it is fairly obvious that mergers are not the primary cause of AGN triggering, as they do not account for 90\% of the AGNs in our sample. Nevertheless, there seems to be a higher prevalence of advanced mergers in the AGN sample. This is confirmed by performing $\chi^{2}$-tests considering only advanced mergers, pairs and early mergers, and the combination; for which the results can be found in Table~\ref{tab:int_chisq}, including the effect size (calculated as the $\varphi$ coefficient), which measures how strong is the correlation between the categorical variables. Considering a typical significance level of 0.05, the p-values obtained indicate that the difference in advanced mergers fraction is statistically significant, while for pairs and early mergers is not. In all cases we have $\varphi$ values below 0.2, which indicates a weak effect size. 

We also derived the merger excess for each case, computed as the merger fraction in AGNs divided by that in non-AGN controls. These excess values we obtained are highly consistent with those found by \citet{LaMarca2024} and at higher redshift by \citet{Duan2024}. On the other hand, \citet{Kaviraj2019} found no merger excess in their dwarf galaxy AGN sample when compared to a control non-AGN sample, by selecting AGNs based on WISE infrared photometry. Similar results are reported by \citet{Bichanga2024}, who identified AGNs via SED fitting. Although the origin of these discrepancies is not entirely clear, it should be noted that the sample sizes of \citet{Kaviraj2019} and \citet{Bichanga2024} are considerably smaller ($\text{N}<900$) than those of \citet{LaMarca2024} ($\text{N}=69140$ for the lowest redshift bin), \citet{Duan2024} ($\text{N}=3330$), and this work. Additionally, considering the effect sizes we found, our results are not incompatible with those of \citet{Kaviraj2019} and \citet{Bichanga2024}.

It is worth noting that the interaction fractions of the respective AGN samples vary considerably among these works. This is possibly in part due to the differences in stellar mass distribution range, and most likely due to the different AGN identification techniques, since the amount of detected AGNs varies considerably depending on the selected method \citep[e.g., see][]{CalderonCastillo2024}. 

More importantly, in the majority of cases, merger fractions in the AGN samples tend to be (well) below 0.5, indicating that most AGNs in dwarf galaxies are not triggered by merger events, meaning that secular processes are a more likely channel for AGN ignition. This is analogous to the seemingly contradicting results in the massive galaxy regime. While it has widely been observed that the AGN fraction increases in interacting massive galaxies \citep[e.g.,][]{Ellison2011, Ellison2013, Satyapal2014, Steffen2023, Barrows2023, Li2023}, the merger/interacting galaxy fraction of AGN samples based on different detection methods has been observed to be below 0.5 in most cases, and with modest to non-significant merger excess \citep{Villforth2023}.

\begin{table}
    \centering
        \caption{Obtained p-values and effect sizes from the $\chi^{2}$-tests and merger excess for different merger groups.}
    \label{tab:int_chisq}
    \begin{tabular}{lccc}
        \hline
        Group & $p_{\chi^{2}}$ & $\varphi$ & Merger excess \\
        \hline
        (Full Sample) \\
        Advanced mergers & $3 \times 10^{-7}$ & 0.09 & $3.31 \pm 0.83$ \\
        Early mergers/pairs & 0.361 & 0.02 & $1.20 \pm 0.22$ \\
        All mergers/pairs & $1.6 \times 10^{-5}$ & 0.08 & $1.83 \pm 0.26$ \\       
        \hline
        (Without SF-AGN) \\
        Advanced mergers & 0.061 & 0.04 & $1.81 \pm 0.59$ \\
        Early mergers/pairs & 0.054 & 0.04 & $1.58 \pm 0.34$ \\
        All mergers/pairs & 0.005 & 0.06 & $1.65 \pm 0.29$ \\       
        \hline
    \end{tabular}
    \tablefoot{The effect sizes are measured by the $\varphi$ coefficient.}
\end{table}

\subsection{The role of the environment} 
\begin{figure}
    \includegraphics[width=\columnwidth]{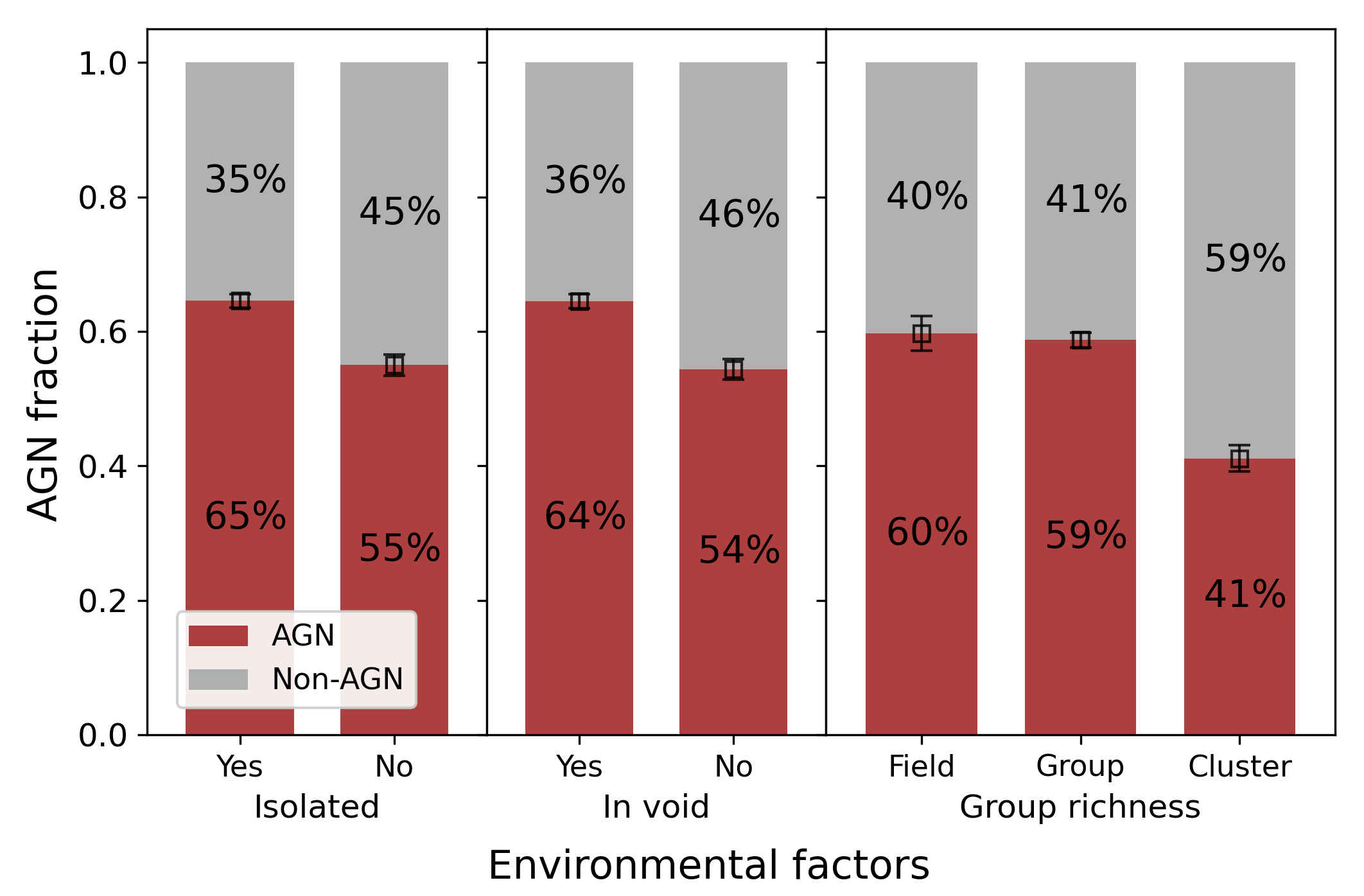}
    \caption{AGN fraction for the three environmental factors used in this work. In all three cases, the AGN fraction decreases with higher population density conditions.}
    \label{fig:environ_AGN}
\end{figure}

We now investigate the impact of the environmental factors on the presence of AGNs in dwarf galaxies. Figure~\ref{fig:environ_AGN} shows the AGN fraction of our sample for galaxies that are isolated, in voids and part of a group/cluster. We find a consistent trend in all three factors: the AGN fraction tends to be higher for galaxies located in less dense environments. We confirm this by performing $\chi^{2}$-tests to determine the statistical differences in the AGN fractions of the samples of each factor, as shown in Table~\ref{tab:env_chi_test}. Since the group richness factor has more than two categories, its effect size is measured by the Cram{\'e}r $V$ instead of the $\varphi$ coefficient. Except for field vs. group galaxies, all differences are statistically significant, although with a weak effect size ($\varphi, V < 0.2$). These results indicate that being part of a cluster has the strongest (environmental) effect on dwarf galaxy quenching. This is in agreement with \citet{Whitherspoon2024}, who identified AGNs by applying the [NII]-BPT and [SII]-BPT diagrams on a sample also extracted from MaNGA (although using a stellar mass limit of $M_{\star}\leq 10^{9.7}~\text{M}_{\odot}$) and found that low-mass galaxies hosting AGNs are more likely to be found in isolation or in low-mass groups in comparison to non-AGN galaxies. The results are also consistent with \citet{Constantin2008}, who found that AGNs are more common in voids than in walls for a sample of galaxies with stellar mass $M_{\star}\leq 10^{10.5}~\text{M}_{\odot}$ and r-band absolute magnitude $M_{r} < -20$.

Simulations also agree with our results. For instance, using \textsc{IllustrisTNG}, \citet{Kristensen2021} found a preference for higher density environments for non-AGN dwarf galaxies. Similar results were reported by \citet{Tremmel2024} using the \textsc{Romulus} cosmological simulation, who found that dwarf galaxies in clusters are less likely to host X-ray luminous AGNs than those in the field.

In contrast, several observational works report no significant difference between the environments of AGN and non-AGN dwarf galaxies. For example, \citet{Kristensen2020} identified AGNs based on emission-line fluxes in a sample of 62258 dwarf galaxies and did not find any environmental difference between AGN and non-AGN dwarf galaxies in the local Universe. Similar results were reported by \citet{Manzano2020}, and also by other authors who identified AGNs based on radio emissions \citep{Davis2022}, SED fitting \citep{Bichanga2024} or even combinations of methods at different wavelengths \citep{Siudek2022}. Particularly for voids, \citet{Liu2015} report that AGNs in voids are similarly abundant to those in walls. 

Interestingly, some authors have found that the AGN fraction actually increases in higher density environments using samples that include but are not limited to dwarf galaxies. For instance, \citet{Peluso2022} observed a significantly higher AGN incidence in galaxies undergoing ram pressure stripping, which happens most efficiently in clusters and massive groups \citep{Hester2006}; and \citet{Hashiguchi2023} found that the AGN fraction of cluster galaxies is always higher than that of field galaxies. However, in both cases it is possible that the results are mainly driven by the non-dwarf galaxy portion of the respective samples. A more dwarf-regime-restricted example is the work of \citet{Amiri2019}, who found that for galaxies with $M_{\star} < 10^{10.2}~\text{M}_{\odot}$, the AGN fraction in clusters is only slightly but systematically higher than in voids.

Similarly to the previous sections, we investigate if the discrepancies between our results and those mentioned above are related to SF contamination, by removing the SF-AGN from our sample. Unlike the case of mergers, we actually find a qualitative difference in the results, as shown in Fig.~\ref{fig:environ_AGN_NoSFAGN}: except for voids, the AGN fraction of dwarf galaxies is now higher for denser environments. A possible explanation for this is that the SFR of dwarf galaxies tends to decrease in denser environments, as suggested by Fig.~\ref{fig:SFR_environment} and observed by other authors (e.g., \citealp{Geha2012, Stierwalt2015} although contradicting results were found by \citealp{Siudek2022}), since as a consequence SF-AGN galaxies are more likely to be concentrated in the lower density environments than composite and robust AGN galaxies. This drastic change could also explain the differences between the results shown in Fig.~\ref{fig:environ_AGN} and the aforementioned works since, except for \citet{Peluso2022} and \citet{Whitherspoon2024} (the results of the latter being consistent with ours), none of them use IFU spectroscopy, which has the advantage of being able to reveal the presence of AGNs that may be missed by other methods \citep{Mezcua2020, Mezcua2024}. The discrepancy between our work and \citet{Peluso2022} is possibly due to the fact that they only used the [NII]-BPT diagram to identify AGNs, which in our case would result in all SF-AGN being classified as just SF \citep[see Sect.~\ref{sec:AGNsel} and][]{Mezcua2024}.

\begin{table}
    \centering
        \caption{Obtained p-values and effect sizes obtained from the ${\chi}^{2}$-tests for each environmental factor.}
    \label{tab:env_chi_test}
    \begin{tabular}{lcccc}
        \hline
          Factor & $p_{\chi^{2}}$ & $p_{\chi^{2}\text{-post hoc}}$ & $\varphi$ & $V$ \\
        \hline
        Isolated/Non & $1 \times 10^{-6}$ & - & 0.08 & - \\
        In Void/Non & $2 \times 10^{-7}$ & - & 0.09 & - \\
        Field/Group & - & 0.68 & - & \multirow{3}{*}{0.12} \\
        Field/Cluster & - & $2 \times 10^{-10}$ & - & \\
        Group/Cluster & - & $2 \times 10^{-7}$ & - & \\
        \hline
    \end{tabular}
    \tablefoot{The effect sizes are measured by the $\varphi$ coefficient for the cases with two categories, and by the Cram{\'e}r $V$ coefficient for the case with more than two categories. Field, Group and Cluster share the same value for $V$. The post hoc p-values were derived using the Benjamini-Hochberg correction.}
\end{table}

\section{Summary and conclusions} 
\label{sec:Conclusions}
We used a sample of 3280 dwarf galaxies from the MaNGA survey that were classified by \citet{Mezcua2024} as hosting (or not) an AGN based on three emission line diagnostic diagrams (the [NII-, [SII]-, and [OI]-BPT) and the WHAN diagram. We arranged the sample into a six-stage merger sequence from earlier to later stages based on visual inspection, projected separation ($r_{\text{p}} \leq 50$ kpc), and velocity difference ($|\Delta v|\leq 300~\text{km~s}^{-1}$) in order to assess if dwarf-dwarf mergers trigger AGNs by measuring the AGN fraction throughout the merger process and comparing it to a control group of non-merger and non-paired galaxies matched in stellar mass, redshift, being isolated, and being in a void. We also studied the contribution of mergers on AGN triggering by measuring the merger fraction in the AGN and non-AGN subsets. In addition, we investigated how the AGN fraction is affected by three environmental factors: isolation, voids, and group richness. With the caveat that red dwarf galaxies may be underrepresented in our sample, our main findings are summarized as follows:
\begin{enumerate}
    \item We observed an increase in the AGN fraction of paired and merging galaxies at separations below 20 kpc. This increase is higher at smaller separations, reaching an AGN fraction of 85\% (which is 1.41 times higher than the respective control group bin) at $r_{\text{p}}<10$ kpc. The AGN fraction peaks at stages 2 (i.e., after the first pass) and 3 (i.e., two distinct nuclei within a common envelope) and remains enhanced until the latest merger stage (remnant).
    This shows that dwarf-dwarf interactions can ignite AGNs independently of being isolated or in a void.
    
    \item There is a significant boost on the AGN fraction of stage 2 mergers, suggesting that the triggering of AGNs occurs mainly from the first pass to posterior interactions.
    
    \item Compared to its non-AGN counterpart, the AGN sample shows a merger excess of 3.31 for advanced mergers, 1.2 for early mergers and pairs, and 1.83 for all mergers and pairs. The differences in the fractions of advanced mergers and all mergers and pairs between the AGN and non-AGN samples are statistically (very) significant but have a weak effect size.
    \item Even though dwarf galaxy mergers can trigger AGNs, the majority of AGNs are not ignited by merger events, leaving secular processes as the more likely AGN triggering mechanism.
    \item The observed merger-AGN connection is very unlikely to be caused by SF contamination, as it remains qualitatively similar when removing all SF-AGN galaxies.
    \item AGN activity in dwarf galaxies seems to be less prevalent in higher density environments, especially in galaxy clusters. This trend, although statistically significant, has a weak effect size for all three environmental factors. 
    \item Since the environment likely has an effect on the SFR, the method and criteria chosen for identifying and selecting AGNs can be critical to determining the relation between AGNs and the environment of dwarf galaxies.
    
\end{enumerate}
Further complementary studies on the AGN fraction in dwarf galaxy mergers using a more extended survey, such as DESI, which contains dwarf AGN candidates at 2.5-5 magnitudes fainter and ten times farther in redshift than those detected in SDSS \citep{Pucha2024}, will provide more complete and statistically robust results. Moreover, surveys with a greater depth would greatly help in reducing or even preventing population biases, such as the one described in Sect.~\ref{sec:low_mass_end}. At the same time, the vast amount of available galaxies in DESI would allow these studies to be expanded to investigate the AGN (and dual AGN) prevalence and the activity in both the dwarf and massive galaxy regimes at different redshifts and to consider different factors (e.g., environment, morphology) simultaneously. This will be crucial to obtaining a more complete and precise model of the AGN-merger relation and the influence of the environment on AGN activity. 

In addition, the sample used in this work contains 17 dual AGN candidates (such as the one displayed in Fig.~\ref{fig:BPT_example}). For 12 of these, a detailed analysis including awarded Chandra X-ray observations and VLBI radio observations will be presented in a forthcoming paper.

The results from these upcoming works will not only expand our understanding of galaxy evolution, but will also provide key constraints on the mass of seed BHs. This is necessary to determine the SMBH formation scenario.

\begin{acknowledgements}
The authors thank the anonymous referee for their insightful comments. M.M. and A.E. acknowledge support from the Spanish Ministry of Science and Innovation through the project PID2021-124243NB-C22.
M.S. acknowledges support by the State Research Agency of the Spanish Ministry of Science and Innovation under the grants ``Galaxy Evolution with Artificial Intelligence'' (PGC2018-100852-A-I00) and ``BASALT'' (PID2021-126838NB-I00) and the Polish National Agency for Academic Exchange (Bekker grant BPN/BEK/2021/1/00298/DEC/1). 
HDS acknowledges financial support by the RyC2022-030469-I grant, funded by MCIN/AEI/10.13039/501100011033 and FSE+.
VRM acknowledges support from the Spanish Ministry of Science, Innovation and Universities through the project
PRE2022-104649. This work was partially supported by the program Unidad de Excelencia Mar\'ia de Maeztu CEX2020-001058-M and the European Union's Horizon 2020 Research and Innovation program under the Maria Sklodowska-Curie grant agreement (No. 754510). This work used public data from the Sloan Digital Sky Survey (SDSS), NASA-Sloan Atlas (NSA), Two Micron All Sky Survey (2MASS), Wide-field Infrared Survey Explorer (WISE), Galaxy Evolution Explorer (GALEX) and the NASA/IPAC Extragalactic Database (NED; DOI: 10.26132/NED1) and Infrared Science Archive (IRSA; DOI: 10.26131/IRSA537).
Funding for the SDSS IV has been provided by the Alfred P. Sloan Foundation, the Heising-Simons Foundation, the National Science Foundation, and the Participating Institutions. SDSS acknowledges support and resources from the Center for High-Performance Computing at the University of Utah. SDSS telescopes are located at Apache Point Observatory, funded by the Astrophysical Research Consortium and operated by New Mexico State University, and at Las Campanas Observatory, operated by the Carnegie Institution for Science. The SDSS web site is \url{www.sdss.org}.
SDSS is managed by the Astrophysical Research Consortium for the Participating Institutions of the SDSS Collaboration, including Caltech, The Carnegie Institution for Science, Chilean National Time Allocation Committee (CNTAC) ratified researchers, The Flatiron Institute, the Gotham Participation Group, Harvard University, Heidelberg University, The Johns Hopkins University, L'Ecole polytechnique f\'{e}d\'{e}rale de Lausanne (EPFL), Leibniz-Institut f\"{u}r Astrophysik Potsdam (AIP), Max-Planck-Institut f\"{u}r Astronomie (MPIA Heidelberg), Max-Planck-Institut f\"{u}r Extraterrestrische Physik (MPE), Nanjing University, National Astronomical Observatories of China (NAOC), New Mexico State University, The Ohio State University, Pennsylvania State University, Smithsonian Astrophysical Observatory, Space Telescope Science Institute (STScI), the Stellar Astrophysics Participation Group, Universidad Nacional Aut\'{o}noma de M\'{e}xico, University of Arizona, University of Colorado Boulder, University of Illinois at Urbana-Champaign, University of Toronto, University of Utah, University of Virginia, Yale University, and Yunnan University.
Funding for the NASA-Sloan Atlas has been provided by the NASA Astrophysics Data Analysis Program (08-ADP08-0072) and the NSF (AST-1211644).
The Two Micron All Sky Survey is a joint project of the University of Massachusetts and the Infrared Processing and Analysis Center/California Institute of Technology, funded by the National Aeronautics and Space Administration and the National Science Foundation.
WISE is a joint project of the University of California, Los Angeles, and the Jet Propulsion Laboratory/California Institute of Technology, and NEOWISE, which is a project of the Jet Propulsion Laboratory/California Institute of Technology. WISE and NEOWISE are funded by the National Aeronautics and Space Administration.
The GALEX mission was developed by NASA in cooperation with the Centre National d'Etudes Spatiales of France and the Korean Ministry of Science and Technology.
The NASA/IPAC Extragalactic Database and Infrared Science Archive are funded by the National Aeronautics and Space Administration and operated by the California Institute of Technology.
We also acknowledge the use of images from the Legacy Surveys; full acknowledgments can be found here: \url{http://legacysurvey.org/acknowledgment/}.
This work made use of Astropy (\url{http://www.astropy.org}): a community-developed core Python package and an ecosystem of tools and resources for astronomy \citep{Astropy2022}.
\end{acknowledgements}

\bibliographystyle{aa.bst}
\bibliography{MaNGA_dwarf_mergers_AGN} 

\begin{appendix}

\section{Finer stage classification}
\label{sec:Fine_stages}
To achieve a more objective classification between stages 1-2 and 4-5 we make use of the morphological asymmetry metric $A$ \citep{Conselice2000}. This was done as follows.

\subsection{Stages 1 and 2} 
After performing the visual classification and pair selection procedures, we review all galaxy pairs classified as stages 1 and 2. Those pairs that show unambiguous tidal features (especially bridges and tails) are directly classified as stage 2. In contrast, galaxy pairs that are clearly separated and show no signs of interaction are classified as stage 1. The remaining cases, which are more ambiguous, are classified using an asymmetry threshold of $A=0.35$ above of which pairs are classified as stage 2, while those below it are classified as stage 1. This threshold was obtained empirically by choosing the value that gave the results that best matched the initial classification. We note that, although this value is the same that was used in \citet{Conselice2003} to distinguish mergers, this is likely coincidental since currently we are only considering early mergers. 

The asymmetry values for most of our sample were taken from the MaNGA Visual Morphologies from SDSS and DESI images value added catalog \citep{VazquezMata2022}. However, in those cases where the parameter \texttt{CAS\_flag} = 0 or where there was a reasonable suspicion that the A value is erroneous (e.g., image contamination or values that were extreme or inconsistent with the image), we re-derived A from DESI r-band images using the python code \textsc{Statmorph} \citep{RodriguezGomez2019} and performing the masking manually.

\subsection{Stages 4 and 5} 
\begin{figure}[hbt!]
    \includegraphics[width=\columnwidth]{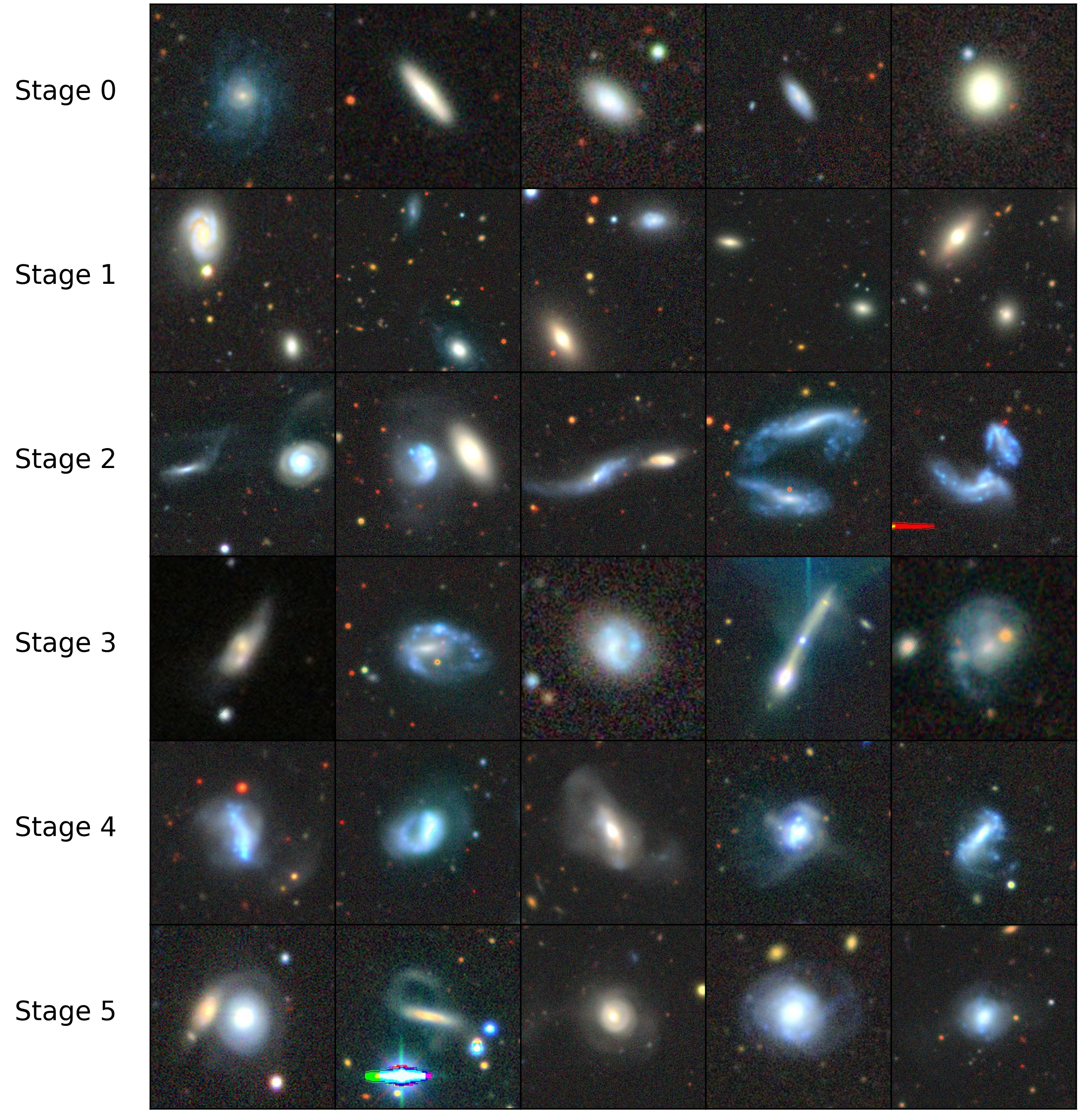}
    \caption{Additional examples of galaxies from our sample at each stage of the merger sequence. All images are extracted from the DESI Legacy Survey, except for the first example of stage 3, which is from SDSS.}
    \label{fig:stages_extended}
\end{figure}

The procedure was similar to that of stages 1 and 2 except for the distinction by tidal features, since they can be present in both stages. The value obtained for the asymmetry threshold was $A=0.21$. All galaxies with values higher than these were labeled as stage 4, while those with lower values were labeled as stage 5. Re-derivation of $A$ was performed with the same criteria and as described for stages 1 and 2. \\
\\
Figure~\ref{fig:stages_extended} shows a sample of five galaxies per merger stage after the final classification. This image is complementary to Fig.~\ref{fig:merger_sequence}.

\FloatBarrier

\section{Estimation of non-merging close pairs}
\label{sec:pairs_to_mergers}
A straightforward method to determine if a galaxy pair is likely to end up merging or not, is to compare the radial velocity difference $|\Delta v|$ of the galaxies to their corresponding radial escape velocity $v_{esc,1D}$, which we estimated following \citet{ArgudoFernandez2014} but assuming the interacting masses are comparable:
\begin{equation} 
    v_{esc,1D} = \sqrt{\frac{2G(\mathcal{M}_{M}+\mathcal{M}_{c})}{3\frac{\sqrt{3}}{\sqrt{2}} r_{p}}},
    \label{eq:logmr_magdiff}
\end{equation}
where $G$ is the universal gravitational constant, $r_{p}$ is the projected separation and $\mathcal{M}_{M}$ and $\mathcal{M}_{c}$ are the dynamical masses of the MaNGA and companion galaxies respectively, which were estimated from their stellar masses assuming $z = 0$ \citep{Moster2013}. The 3 and $\frac{\sqrt{3}}{\sqrt{2}}$ factors in the denominator relate the relative radial velocity (assuming an isotropic distribution) and projected separation to their 3D physical counterparts, respectively. This results in $\sim87\%$ of close pairs that will end up as mergers.

Alternatively, we can use a more empirical method presented by \citet{Ventou2019}, who used the \textsc{Illustris-1} simulation to derive a formula to compute the merging probability of a close pair as a function of the radial velocity difference and projected separation: 
\begin{equation} 
    W(r_{p},|\Delta v|) = 1.407 \pm 0.035 \space e^{-0.017 \pm 0.0004 \space r_{p} - 0.005\pm0.0001 \space |\Delta v|},
    \label{eq:logmr_magdiff}
\end{equation}
which can be applied to each pair to then derive the merging fraction of all close pairs. From this we obtain that $\sim48\%$ of the close pairs will merge. We note that there are another two formulas which are more specific for stellar masses above and below $M_{\star} = 10^{9.5}~\text{M}_{\odot}$ respectively, but the results differ very little. Thus, we expect that the portion of close pairs (stage 1) that will not end up merging is between $\sim13\%$ and $\sim52\%$.

\section{Photometric data for the SED fitting} 
\label{App_Photometry}
In order to ensure the SED fitting quality for the galaxies that were not properly deblended or have no available stellar mass in the NSA catalog (see Sect.~\ref{sec:mass_ratio}), we required as complete and reliable photometric data as available for our main sample and companion galaxies. The data collecting procedure for the different photometric bands used is detailed below.

\subsection{SDSS}
We used the SDSS CasJobs tool to find all photometric objects within 5\arcsec around each of the affected galaxies. Since in many cases the SDSS DR18 deblending algorithm interprets regions of a galaxy as a separate galaxy (generally with a very dim magnitude) we had to visually inspect all the detected photometric objects to make sure we select the correct one. In a very few cases where there was no adequate object or the photometry was flagged as unreliable, photometric values were inspected and taken from DR7.
To have an optimal balance between the accuracies of color and total luminosity, we followed the same procedure of \citet{Maraston2013} and used the model magnitudes but scaled to the i-band cmodel magnitudes (or Petrosian magnitudes for the DR7 cases).

\subsection{GALEX}
Since many sources of our sample appear in different GALEX-based catalogs, we crossmatched with each one and combined the results by complementing the missing values (or replacing them when the following catalog had more complete values for the galaxy) in the following order: Improved GALEX photometry of $z<0.3$ SDSS galaxies \citep{Osborne2023}, Revised catalog of GALEX UV sources \citep{Bianchi2017}, Catalogs of unique GALEX sources from GALEX fifth data release \citep[GR5;][]{Bianchi2011}, and the GALEX public archive at MAST \citep{Conti2011}. The crossmatch radius used was 8\arcsec, and all matches with a separation greater than 5\arcsec were visually inspected to ensure they were correctly matched.

\subsection{allWISE/2MASS}
The allWISE photometry was obtained by a simple crossmatch. In the case of 2MASS, we first used the XSC and then complemented it with the Point Source Catalog. The crossmatch radius and procedure for both surveys were the same as for GALEX. 

\subsection{Extinction}
Finally, to correct for extinction we took the color excess values $E(B-V)$ of \citet{Schlafly2011} from the NASA/IPAC IRSA Galactic Dust Reddening and Extinction service. We used the extinction coefficients $R_{\lambda}$ from \citep{Zhang2023} for all bands.

\section{Estimation of missing close companions due to flux limits} 
\label{missing_comp_estimation}
\begin{figure}
    \includegraphics[width=\columnwidth]{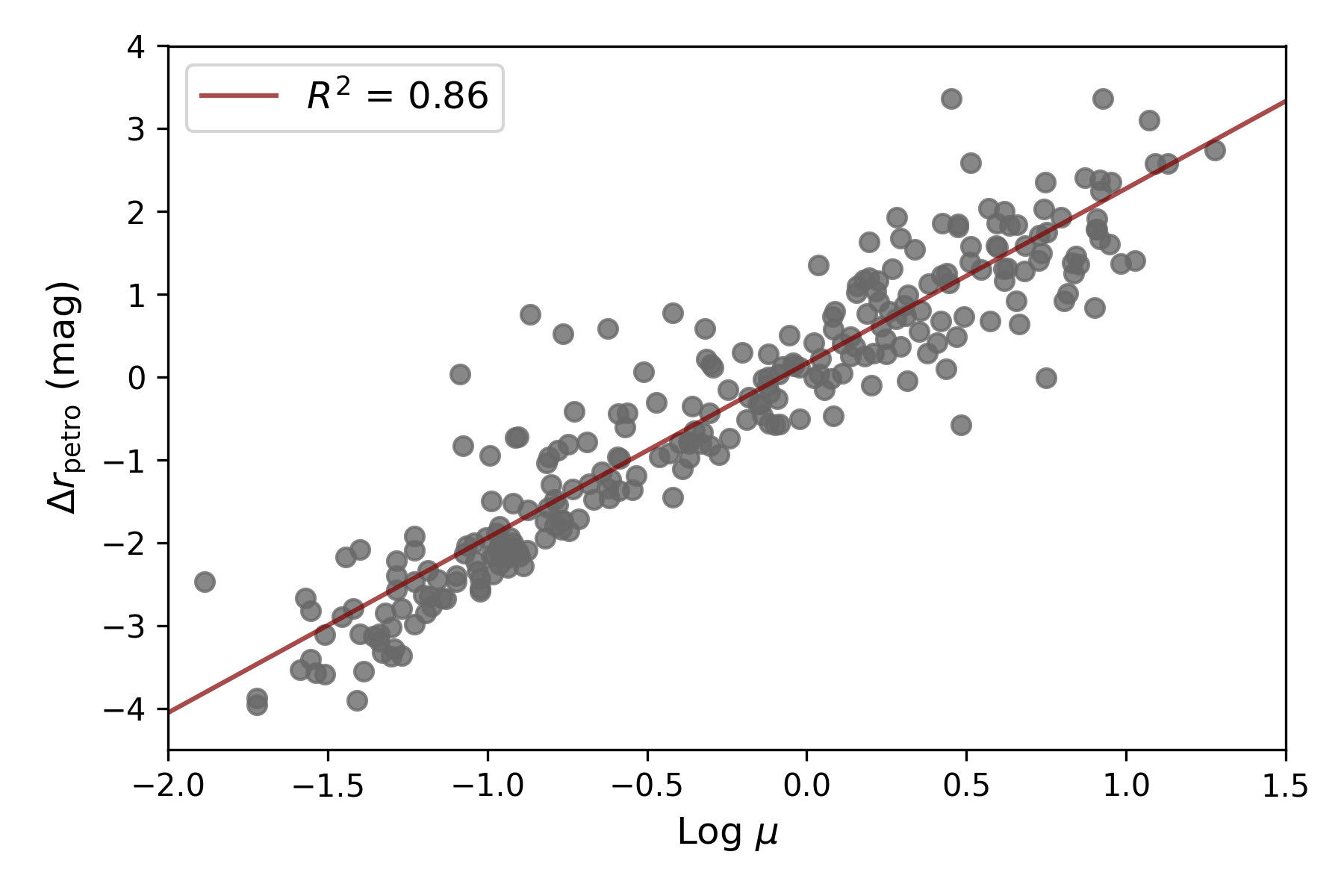}
    \caption{Relationship between the Petrosian r-band magnitude differences of galaxies from the MaNGA dwarf sample and their companions and the logarithm of their respective mass ratio. The red line is the best fitting line to the data, corresponding to Eq.~\ref{eq:logmr_magdiff}.}
    \label{fig:plot_logmr_mag}
\end{figure}

As stated in Sect.~\ref{sec:pair_selection}, all suitable companion galaxies for our sample should have mass ratios $\mu \leq 10$. This should also roughly correspond to a maximum magnitude difference, where the brighter magnitude corresponds to the MaNGA galaxy of the pair. Based on this idea, we used all the close companions we found for our sample (including massive and secondary companions, which were excluded from the main analysis) to derive an empirical correlation between the logarithm of the mass ratio and the difference in the r-band Petrosian magnitude:
\begin{equation} 
	\Delta r = (2.1 \pm 0.05) *\log(\mu) + (0.17 \pm 0.04).
	\label{eq:logmr_magdiff}
\end{equation}
This equation corresponds to the best fitting line based on the Least Squares Method, which yielded a determination coefficient of $R^2 = 0.86$ and is displayed in Fig.~\ref{fig:plot_logmr_mag}. 

From this relation we found that for $\mu = 10$, the maximum allowed magnitude difference is $\Delta r = 2.27$. We then applied this value to the r-band Petrosian magnitude of each galaxy in our MaNGA sample that is not a merger and has no detected companions (i.e., stage 0) to find how many could have a suitable companion that is beyond the detection limit ($r>17.77$). We refer to this group as the limit-affected subset and denote the number of members in it as $N_\text{A}$. Next, we define another subset made of those MaNGA galaxies that have a maximum allowed magnitude difference $\Delta r_\text{max} \leq 17.77$ (i.e., galaxies not affected by the flux limit) to derive the probability of having a suitable companion for our sample $p_{\text{c}}$ as: (number of detected suitable companions of the subset)/(total number of galaxies in the subset). 

\begin{figure}
    \includegraphics[width=\columnwidth]{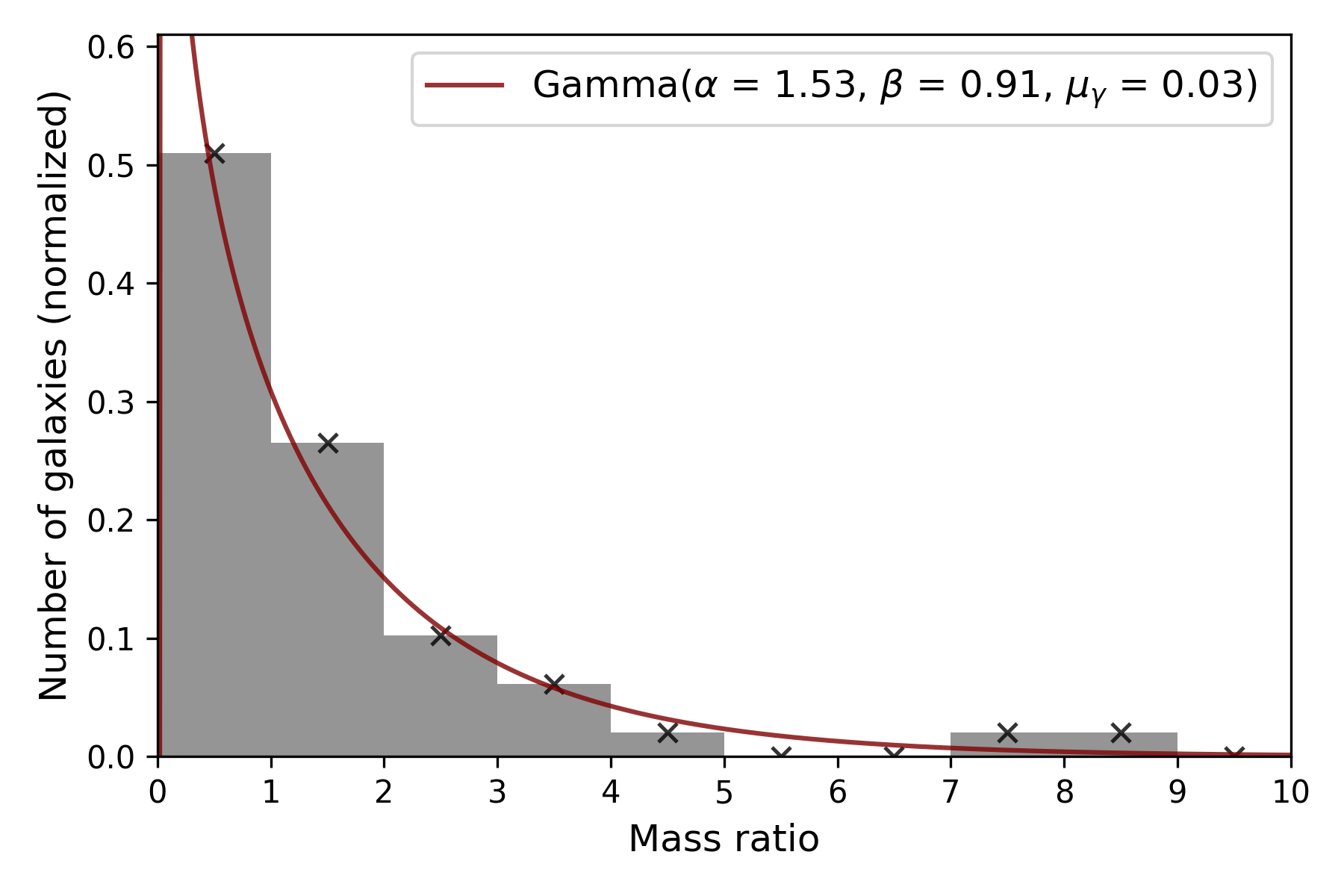}
    \caption{Mass ratio distribution of the galaxy pairs that are unaffected by the flux limit. The \textit{y} axis is normalized so that the sum of all bins is 1. The black crosses show the center of the bin and the red line is the modeled gamma distribution with $\alpha=1.53$, $\beta=0.91$ and $\mu_{\gamma}=0.03$, where $\alpha$, $\beta$ and $\mu_{\gamma}$ are the shape, scale and location parameters, respectively.}
    \label{fig:chisq_fit}
\end{figure}

An important aspect to consider is that most galaxies are only affected partially by the flux limit, depending on their magnitude. For example, the search for a companion of a galaxy with a magnitude of $r = 16.0$ would only be affected by the flux limit for potential companions that yield a mass ratio greater than $\mu=4.6$, which corresponds to the difference between the example galaxy and the SDSS limit according to Eq.~\ref{eq:logmr_magdiff}. This means that the potentially missing pairs correspond only to the portion of potential companions with a mass ratio greater than the corresponding limit. To find this portion, we modeled the mass ratio distribution of all observed suitable pairs as a gamma distribution as shown in Fig.~\ref{fig:chisq_fit}, for which the parameters were determined using the maximum likelihood estimation method. To assess the quality of the model we ran a KS-test between the model and the data, and obtained a p-value of $\sim 0.6$, suggesting an adequate distribution model. The area under the distribution curve within $0 < \mu \leq 10$ is then normalized, so that given a certain mass ratio limit, we can calculate the portion affected by the flux limit (i.e., the area under the curve beyond the maximum observable mass ratio). We compute this portion for every galaxy of the limit-affected subset and obtain the mean value, which we denote as $p_{\mu\text{-lim}}$.
Next, we have to consider that, although comparatively very few, there are galaxies detected at $r > 17.77$, which we should subtract from the final estimation. In order to do this, we use the ratio of the mean completeness values above and below $r = 17.77$ as $p_{\text{s}} = f_{\text{s}}^{(17.77 < r \leq 20)}/f_{\text{s}}^{(r \leq 17.77)}$, where we have taken $r=20$ as the limit since it becomes extremely unlikely that one of our galaxies has a suitable companion at this magnitude.
Finally, we estimate the total number of missing suitable companions as
\begin{equation} 
    N_{\text{m}} = N_\text{A}*p_{\text{c}}*p_{\mu\text{-lim}}*(1-p_{\text{s}}).
    \label{eq:missing_estimation}
\end{equation}
We also did this entire procedure considering a maximum of $\mu=4$ to be able to distinguish the ranges equivalent to those of major and minor mergers. The corresponding results can be found in Table~\ref{tab:missing_comp_tab}.

\section{The effects of removing SF-AGN}
\begin{figure}[hbt!]
    \includegraphics[width=\columnwidth]{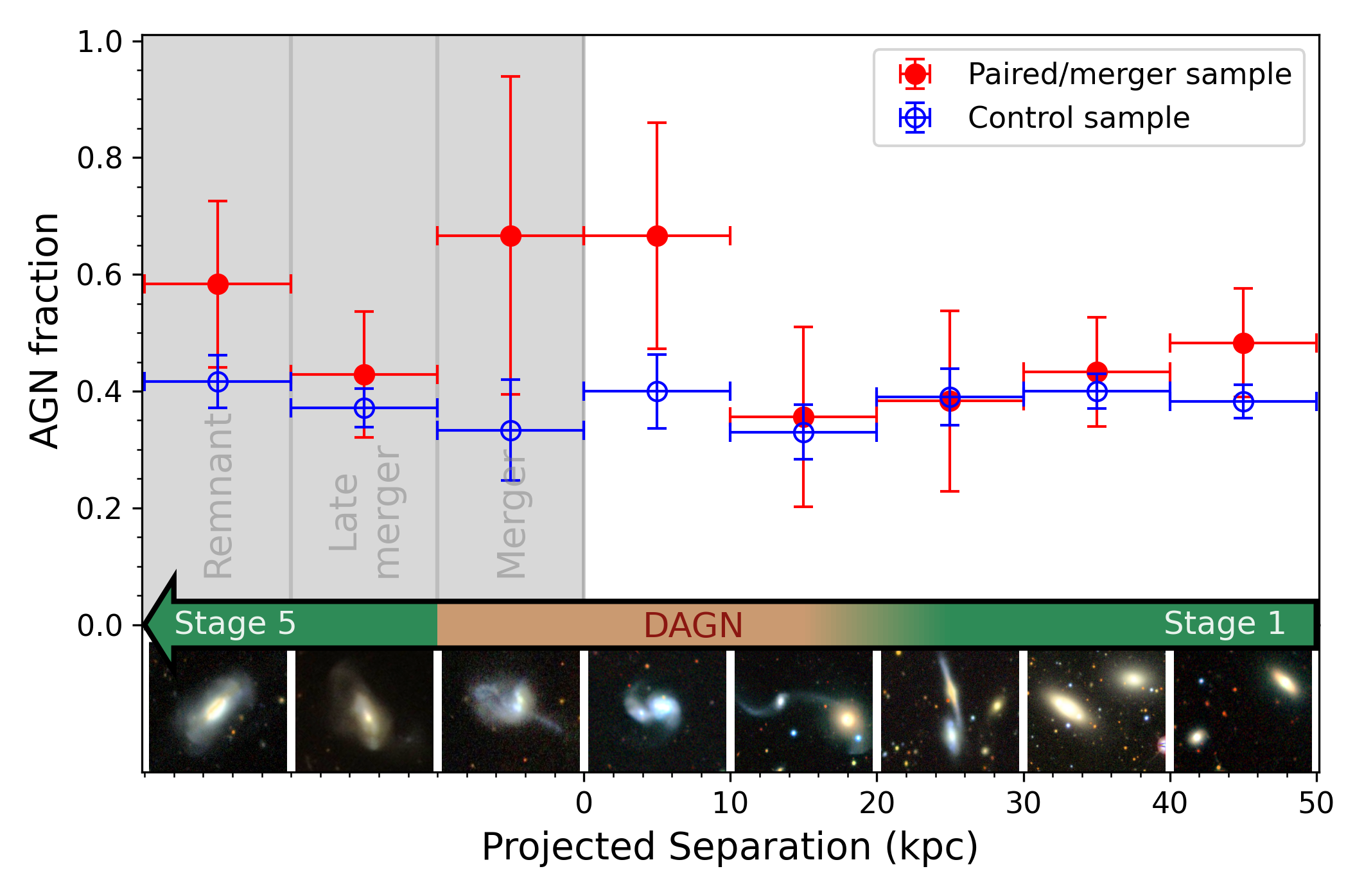}
    \includegraphics[width=\columnwidth]{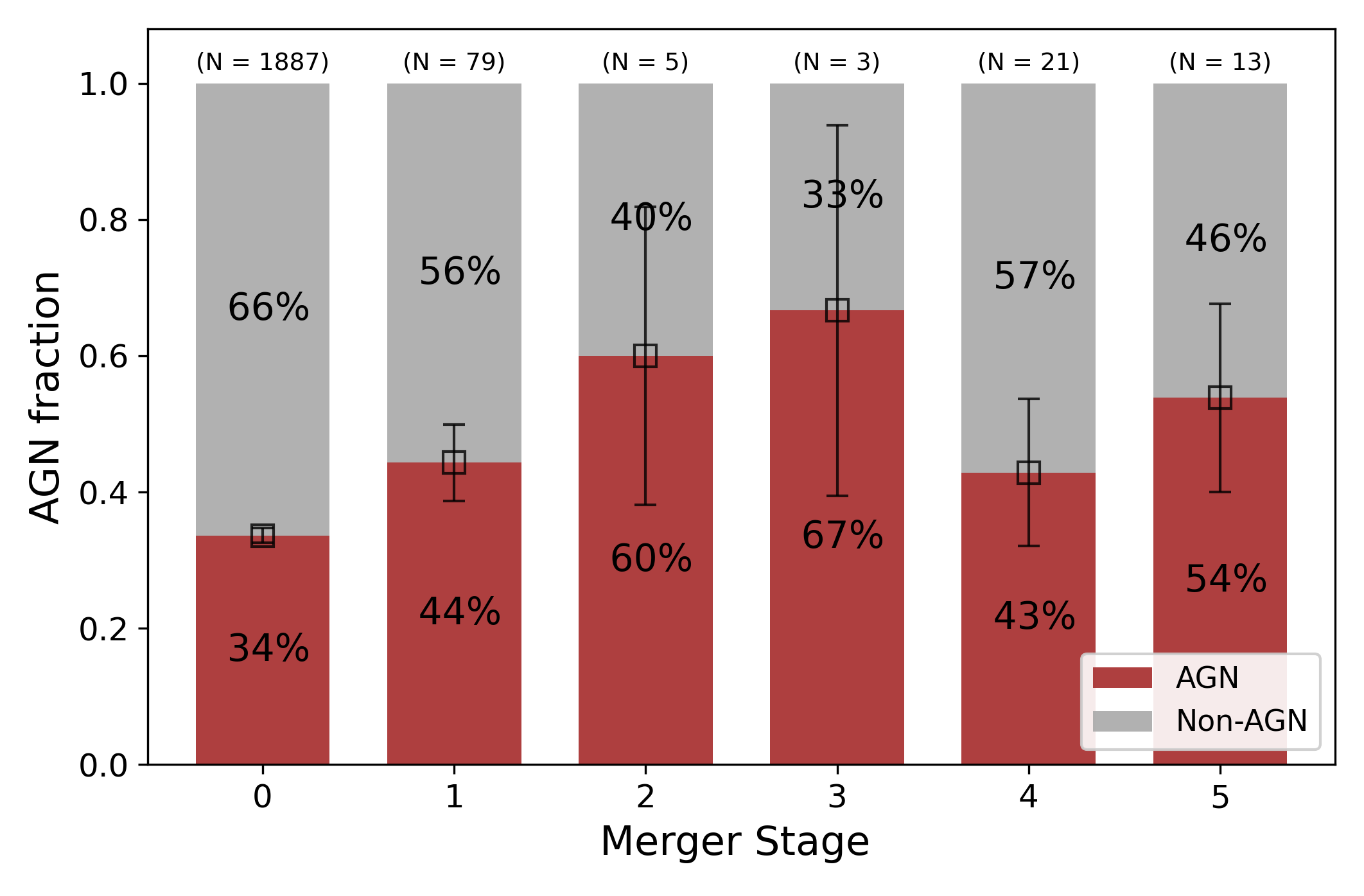}
    \caption{Same as Fig.~\ref{fig:merger_AGNfraction} but after removing all galaxies classified as SF-AGN from our sample.}
    \label{fig:merger_AGNfraction_NoSF-AGN}
\end{figure}

\begin{figure}[hbt!]
    \includegraphics[width=\columnwidth]{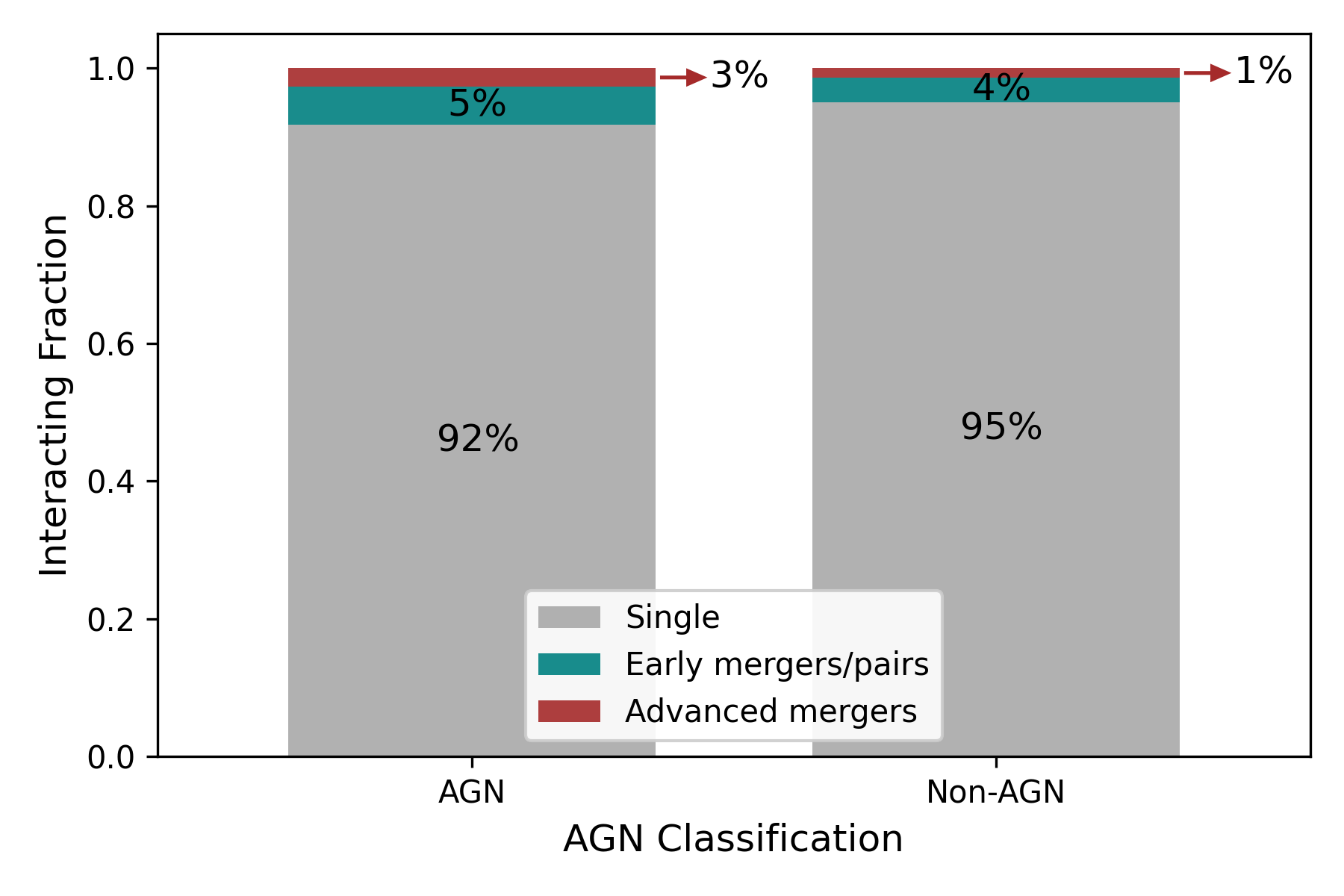}
    \caption{Same as Fig.~\ref{fig:intfrac_AGNnon} but after removing all galaxies classified as SF-AGN from our sample.}
    \label{fig:intfrac_AGNnonNoSFAGN}
\end{figure}

\begin{figure}[hbt!]
    \includegraphics[width=\columnwidth]{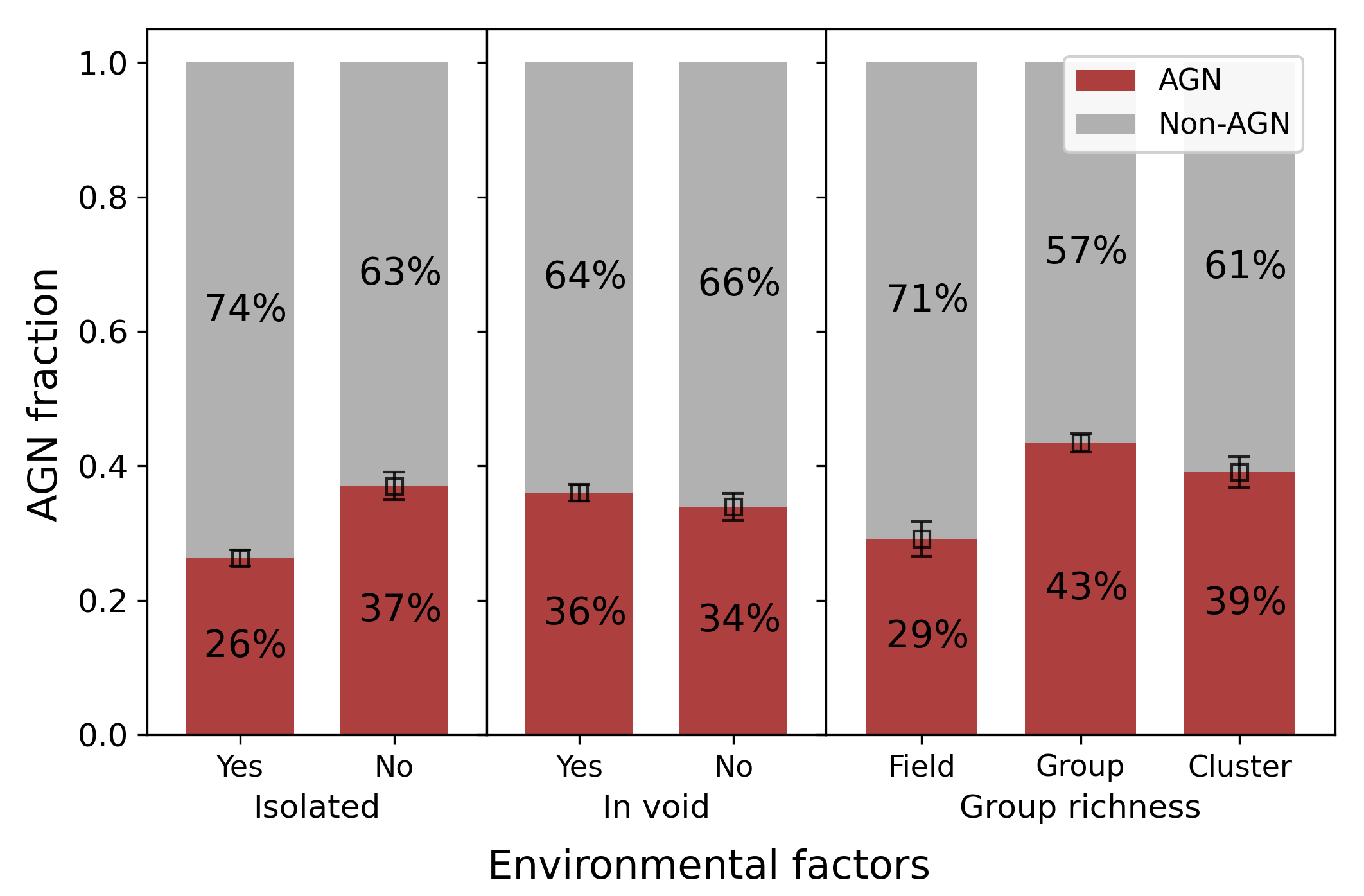}
    \caption{Same as Fig.~\ref{fig:environ_AGN} but after removing all galaxies classified as SF-AGN from our sample.}
    \label{fig:environ_AGN_NoSFAGN}
\end{figure}

Here we show how our results are affected by removing the galaxies classified as SF-AGN from our sample, as discussed in Sect.~\ref{sec:Results}. Figure~\ref{fig:merger_AGNfraction_NoSF-AGN} shows the AGN fraction vs. the projected separation (top panel) and the AGN fraction at each merger stage (bottom panel), Fig.~\ref{fig:intfrac_AGNnonNoSFAGN} shows the fraction of single galaxies, early mergers/pairs and advanced mergers in the AGN and non-AGN subsets, and Fig.~\ref{fig:environ_AGN_NoSFAGN} shows the AGN fraction for the three environmental factors considered in this work.

\begin{figure}[hbt!]
    \includegraphics[width=\columnwidth]{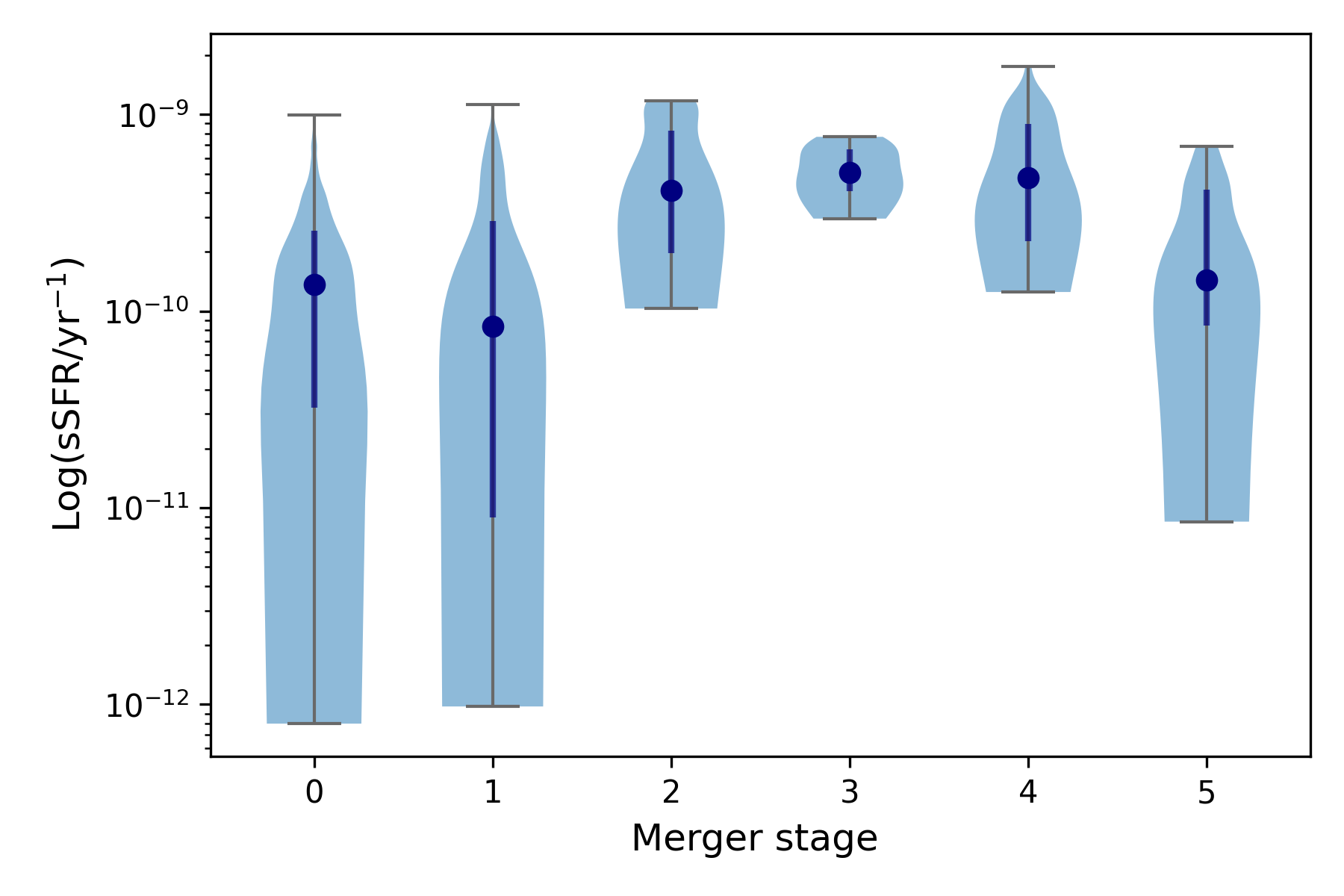}
    \caption{Violin plots of the specific star formation rate throughout the merger sequence. The blue circles indicate the median value, the thick blue line is the interquartile range and the thin gray bars show the distribution limits.}
    \label{fig:SFR_stages}
\end{figure}

Additionally, we show the specific star formation rate (sSFR) distribution of each merger stage for the entire sample in Fig.~\ref{fig:SFR_stages}. As mentioned in Sect.~\ref{sec:AGN_stages}, the lowest sSFR value corresponds to stage 1. Interestingly, this slight decrease of SFR at pre-first pass stages also appears in the works of \citet{CalderonCastillo2024} and \citet{Pan2019} and for mass ranges including dwarf galaxies. However, as in this work, there is not a statistically significant difference in either case.


\begin{figure}[hbt!]
    \includegraphics[width=\columnwidth]{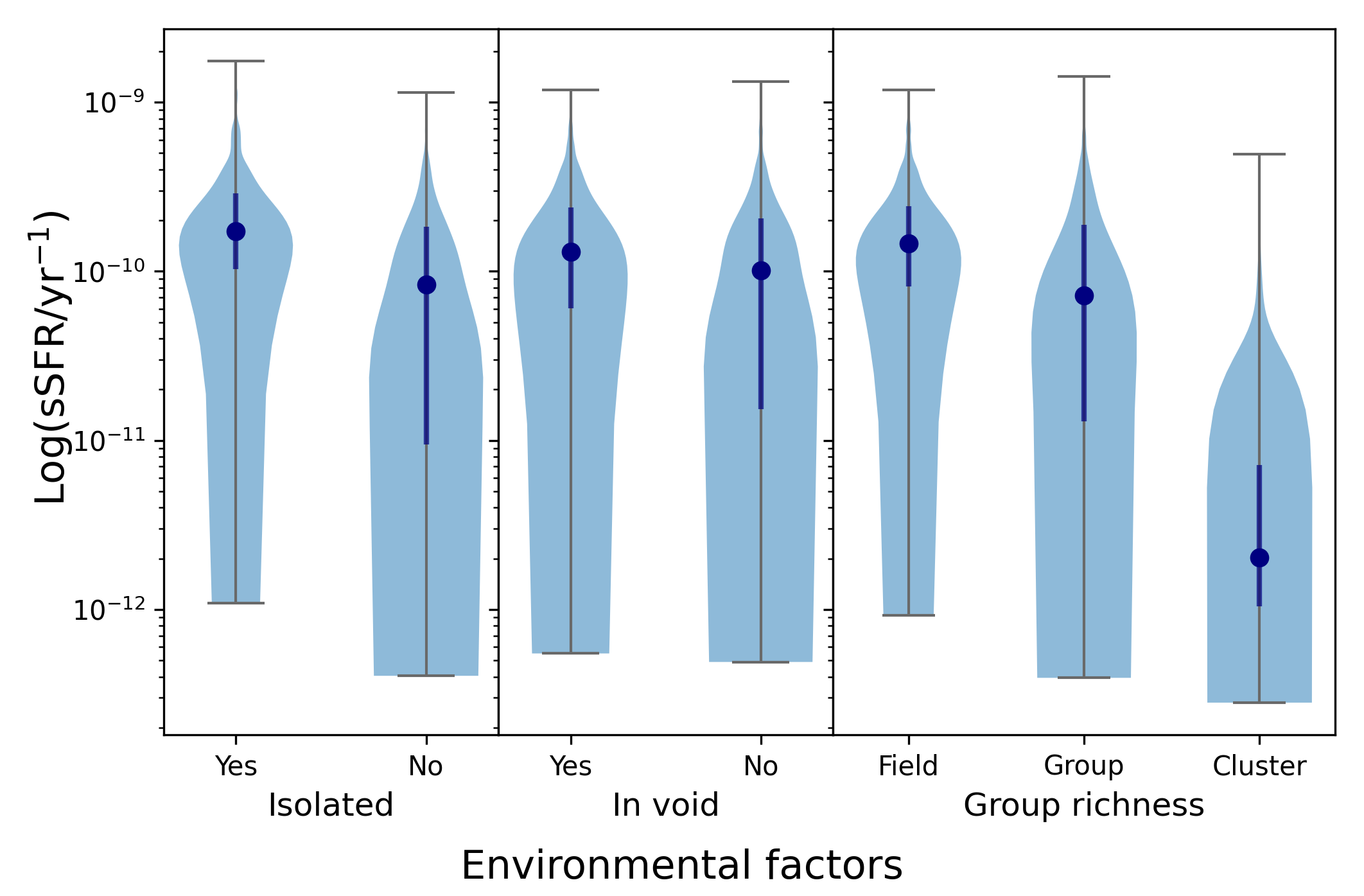}
    \caption{Violin plots of the specific star formation rate for the different environmental factors. The description is the same as in Fig.~\ref{fig:SFR_stages}.}
    \label{fig:SFR_environment}
\end{figure}
\FloatBarrier

\section{Complementary images}
\begin{figure}[!hbt]
    \includegraphics[width=\columnwidth]{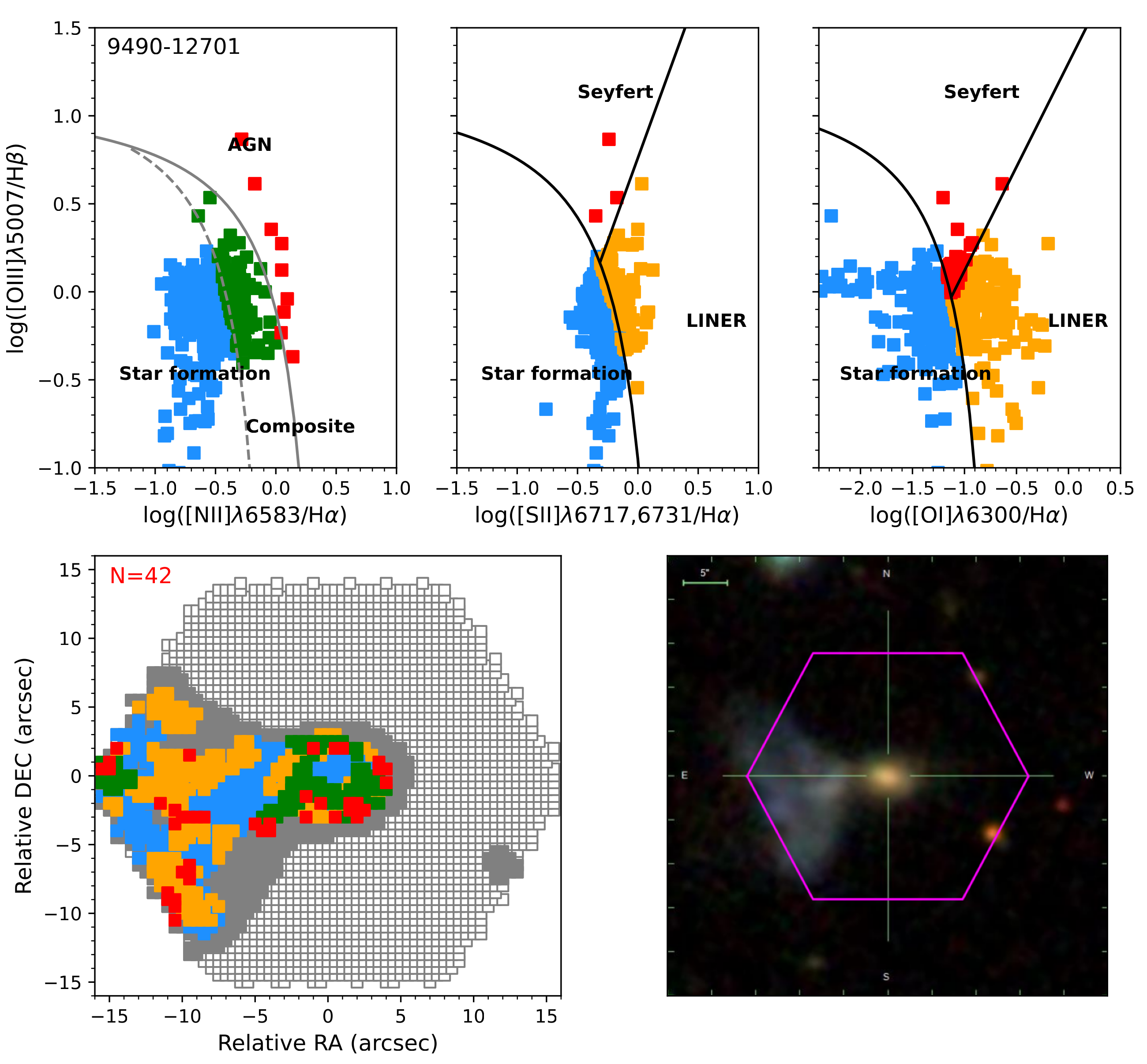}
    \caption{MaNGA analysis of the AGN dwarf galaxy 9490-12701.The description is the same as in Fig.~\ref{fig:BPT_example}.}
    \label{fig:Spaxel_overcount_example}
\end{figure}

Figure~\ref{fig:Spaxel_overcount_example} shows the emission line diagnostic classification of the MaNGA source 9490-12701. The main galaxy (located in the center of the MaNGA field) has less than 20 AGN spaxels, but the corresponding proportion of AGN spaxels is relatively high when considering the small portion of the MaNGA field that the galaxy occupies.

\end{appendix}
\end{document}